\documentclass{aa}
\usepackage[utf8]{inputenc} 

\usepackage{graphicx} 
\usepackage{txfonts} 

\usepackage{xcolor} 
\usepackage{natbib}

\pdfpageattr {/Group << /S /Transparency /I true /CS /DeviceRGB>>}

\bibpunct{(}{)}{;}{a}{}{,} 

\newcommand{\FigureCodeStructure}{%
  \begin{figure}
    \begin{center}
    \includegraphics[width=0.6\linewidth]{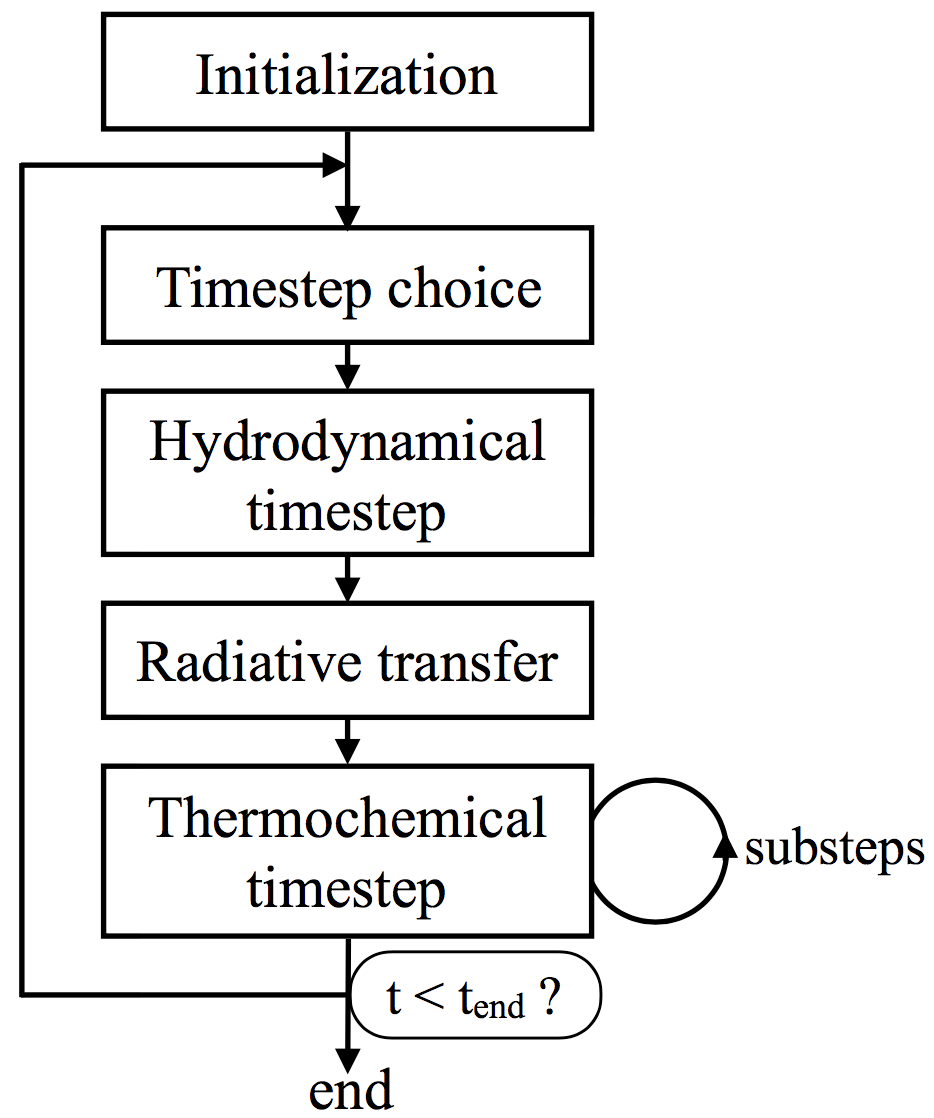}
    \end{center}
    \caption{Code structure.}
    \label{fig:CodeStructure}
  \end{figure}
}

\newcommand{\FigureSodShockTubeTest}{%
  \begin{figure}
    \includegraphics[width=1\linewidth]{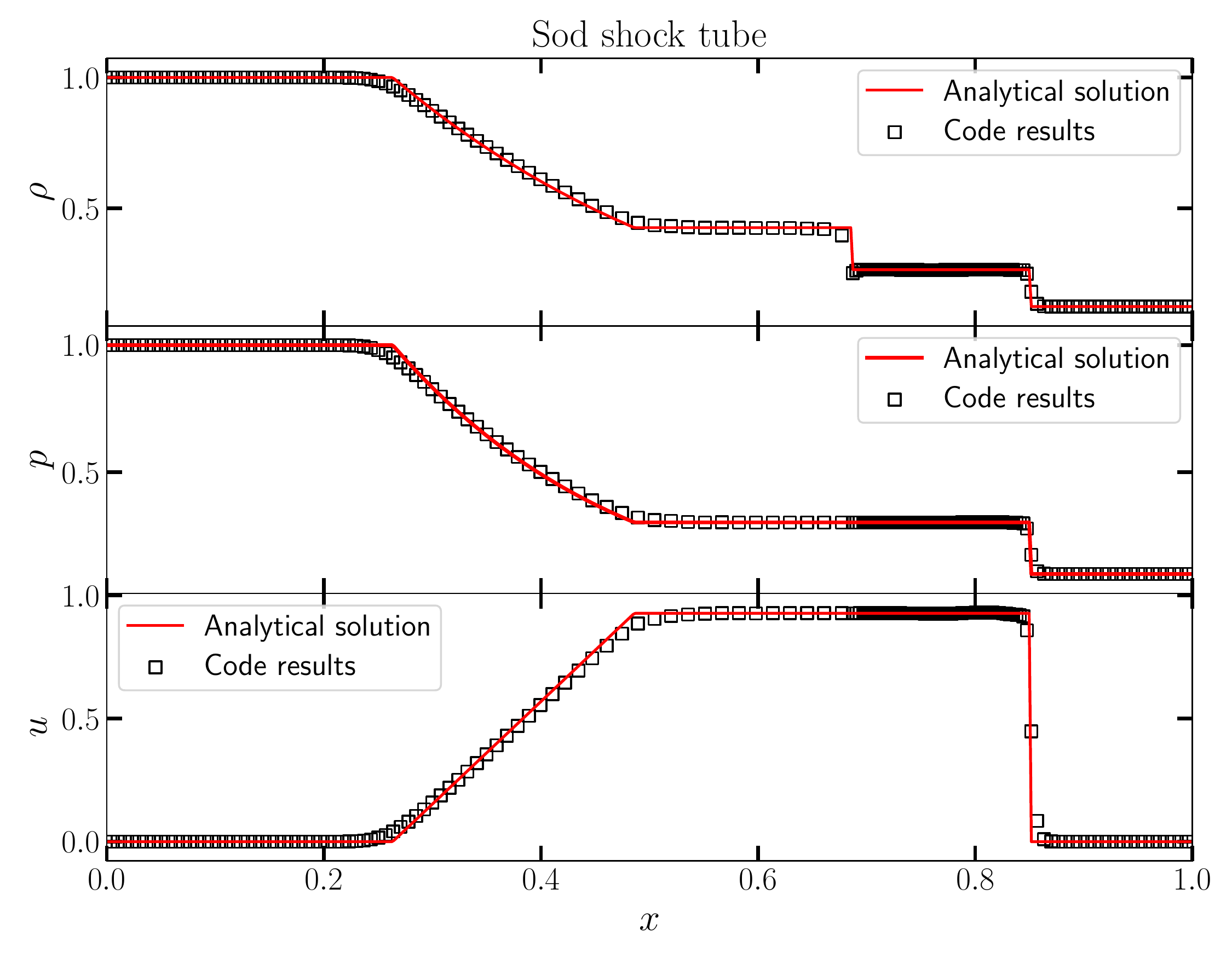}
    \caption{Code verification on the Sod shock tube test problem. The 
    computed mass density $\rho$, pressure $p$, and velocity $u$ are shown as black squares, 
    and compared to the analytical solution (red line).}
    \label{fig:SodShockTubeTest}
  \end{figure}
}

\newcommand{\FigureSedovPlanarBlastWave}{%
  \begin{figure}
    \includegraphics[width=1\linewidth]{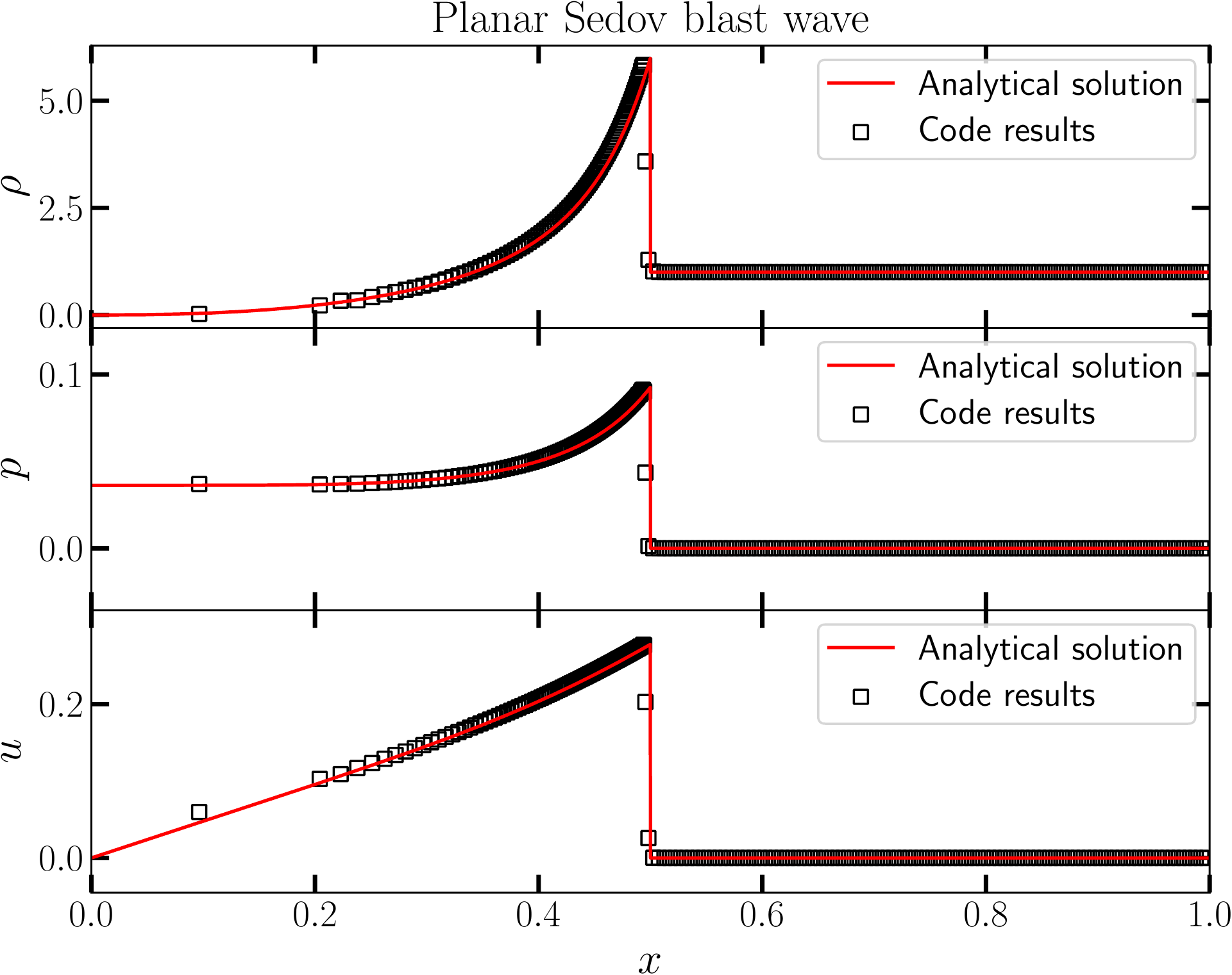}
    \caption{Code verification on the Sedov blast wave test problem in planar geometry. The 
    computed mass density $\rho$, pressure $p$, and velocity $u$ are shown as black squares, 
    and compared to the analytical solution (red line).}
    \label{fig:SedovPlanarBlastWave}
  \end{figure}
}

\newcommand{\FigureSedovSphericalBlastWave}{%
  \begin{figure}
    \includegraphics[width=1\linewidth]{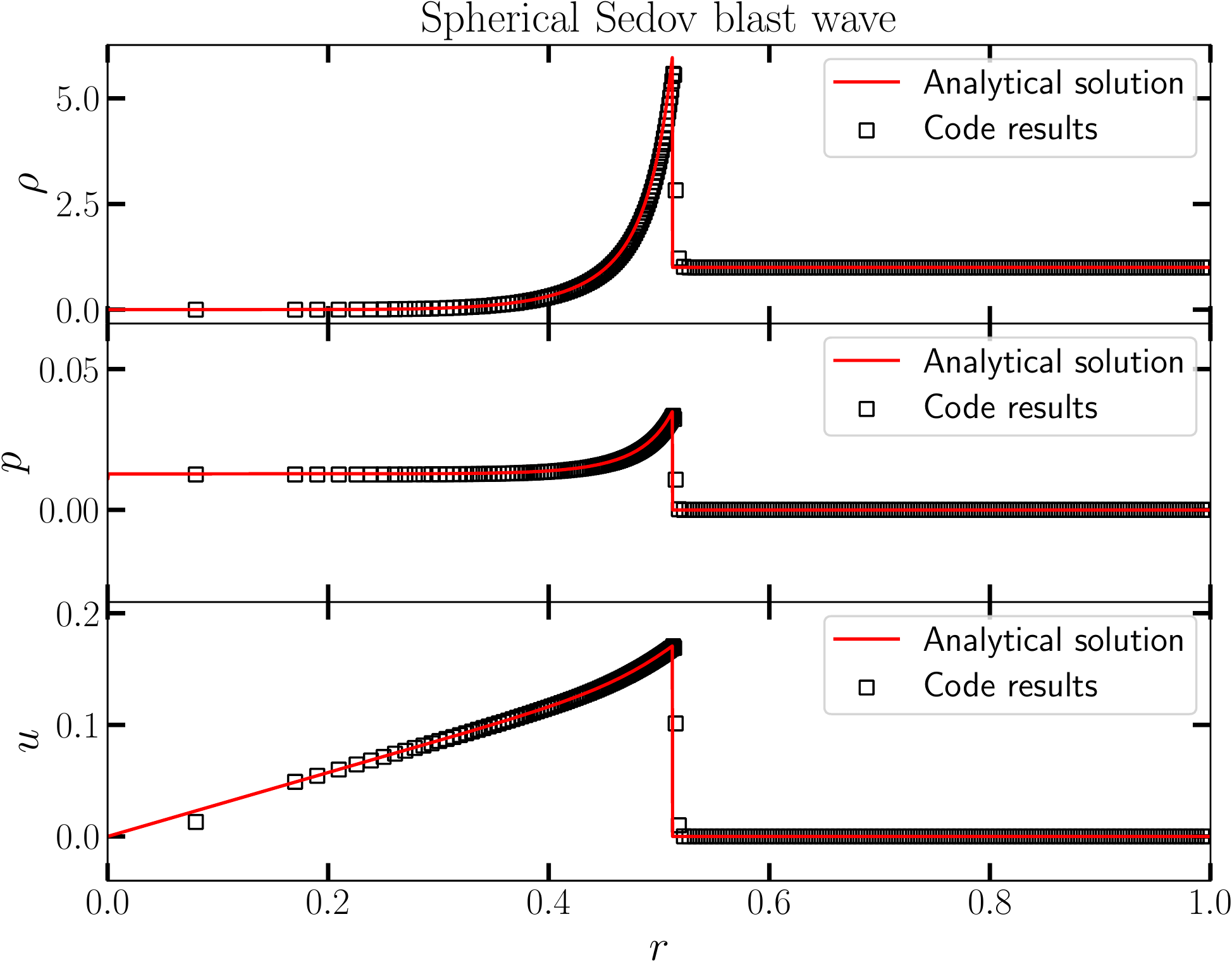}
    \caption{Code verification on the Sedov blast wave test problem in spherical geometry. The 
    computed mass density $\rho$, pressure $p$, and velocity $u$ are shown as black squares, 
    and compared to the analytical solution (red line).}
    \label{fig:SedovSphericalBlastWave}
  \end{figure}
}

\newcommand{\FigureSTARBENCHearlyphase}{%
  \begin{figure}
    \includegraphics[width=1\linewidth]{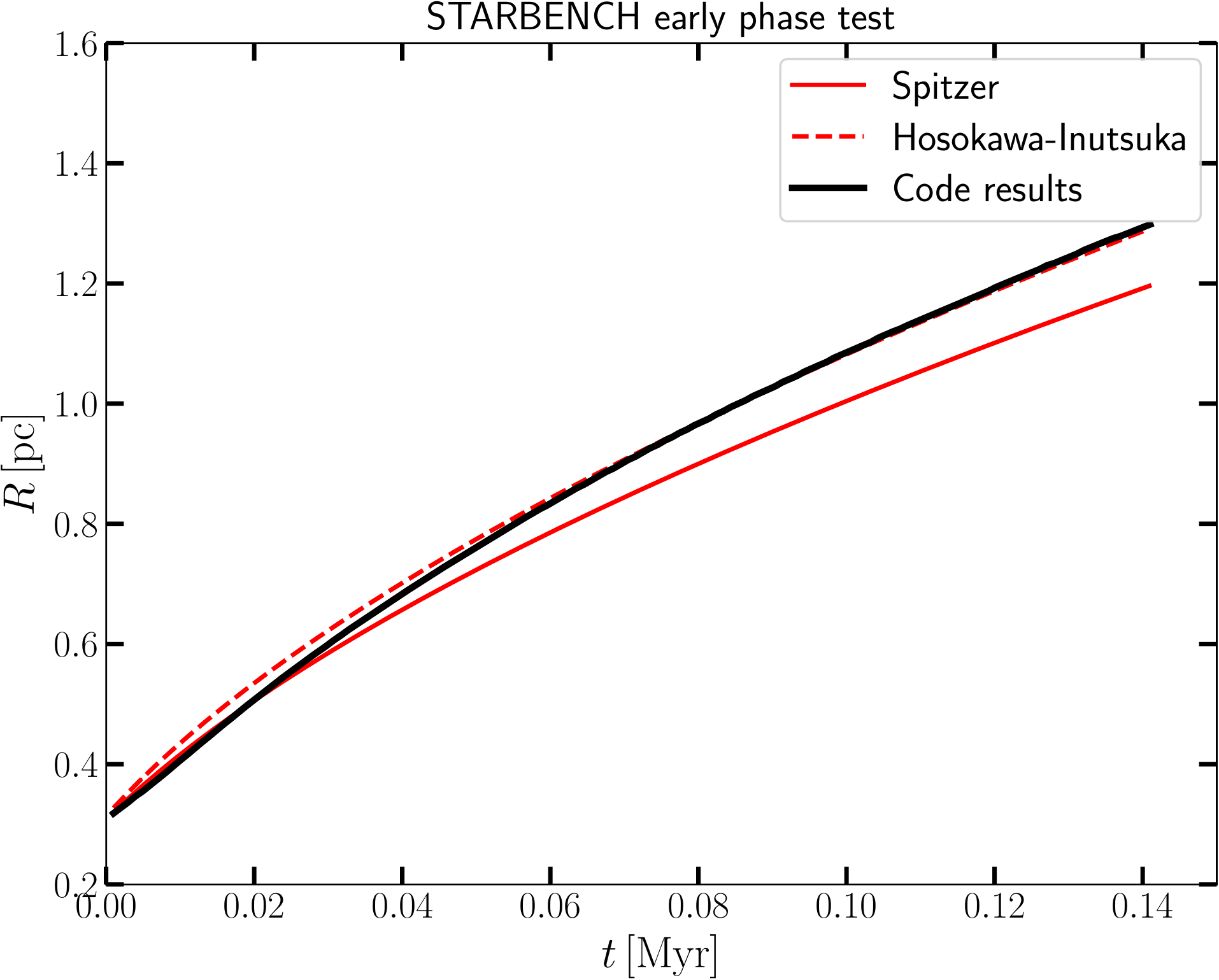}
    \caption{Code verification on the STARBENCH early phase test case, comparing the ionization front
    position during the pressure-driven expansion of an H\textsc{ii} region to various analytical prescriptions.}
    \label{fig:STARBENCHearlyphase}
  \end{figure}
}

\newcommand{\FigureSTARBENCHlatephase}{%
  \begin{figure}
    \includegraphics[width=1\linewidth]{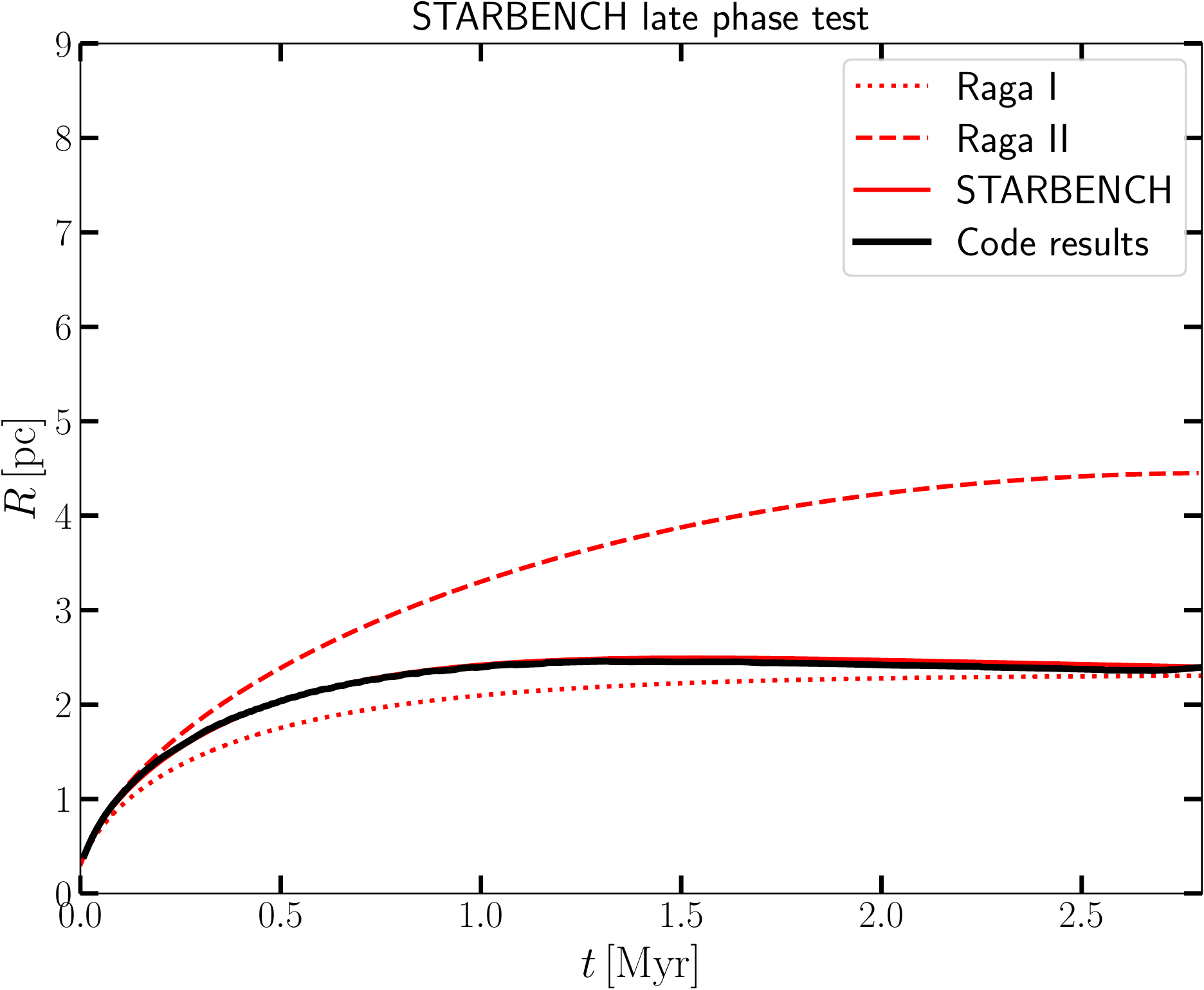}
    \caption{Code verification on the STARBENCH late phase test case, comparing the ionization front
    position during the pressure-driven expansion of an H\textsc{ii} region to various analytical prescriptions.}
    \label{fig:STARBENCHlatephase}
  \end{figure}
}

\newcommand{\FigureEvapDrawing}{%
  \begin{figure}
    \begin{center}
    \includegraphics[width=\linewidth]{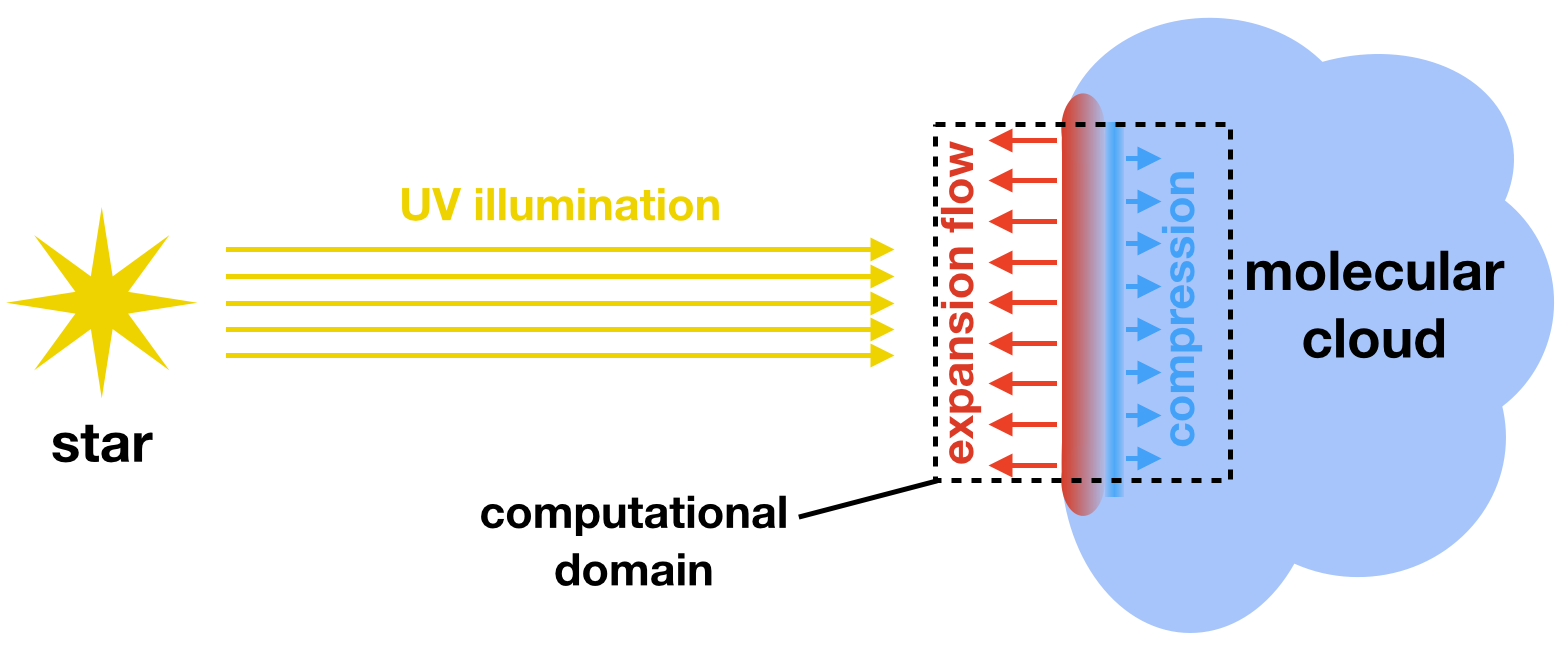}
    \end{center}
    \caption{Schematic view of our photoevaporation scenario.}
    \label{fig:EvapDrawing}
  \end{figure}
}

\newcommand{\FigureExampleRunOne}{%
  \begin{figure*}
    \includegraphics[width=.5\linewidth]{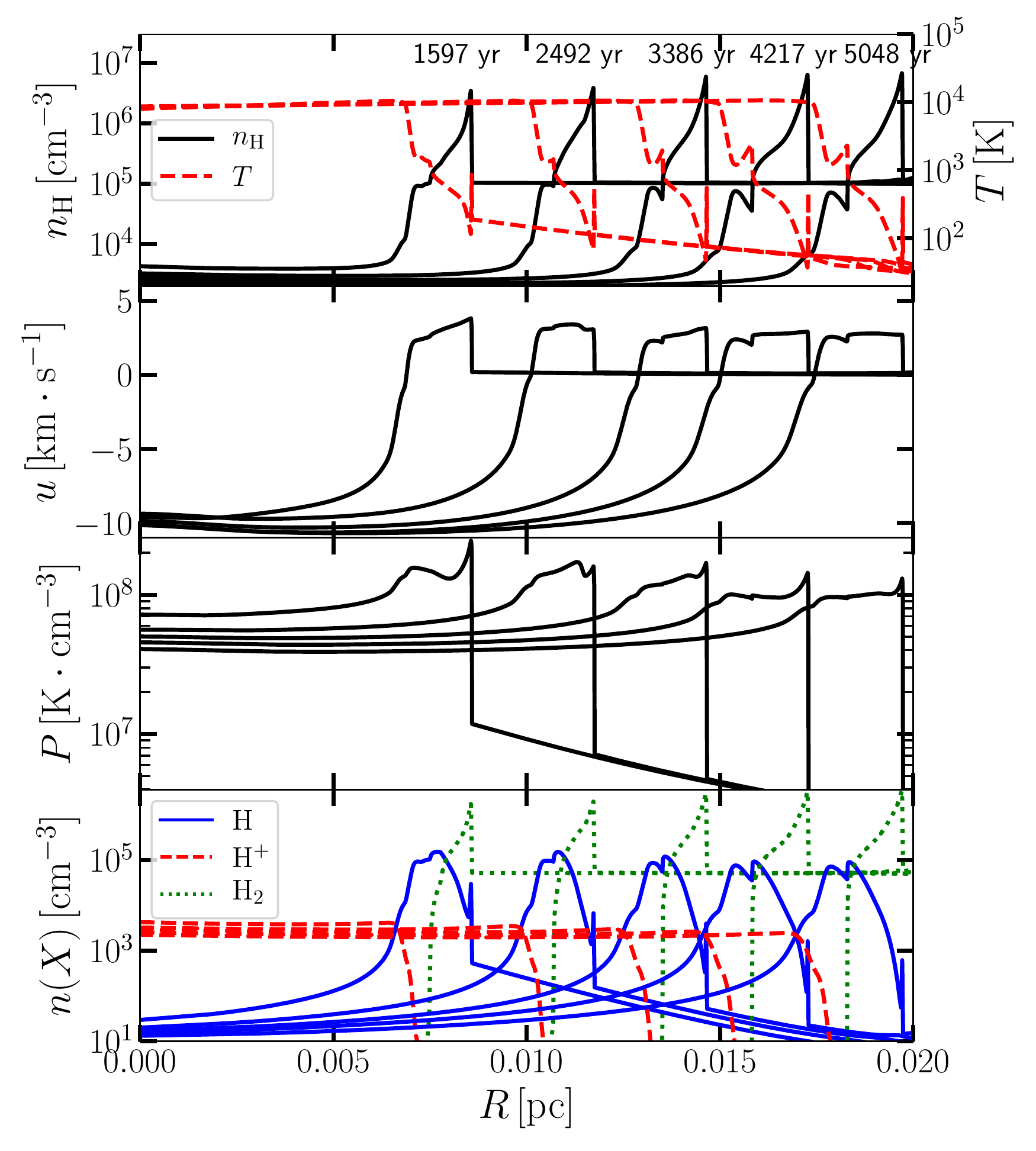} \includegraphics[width=.49\linewidth]{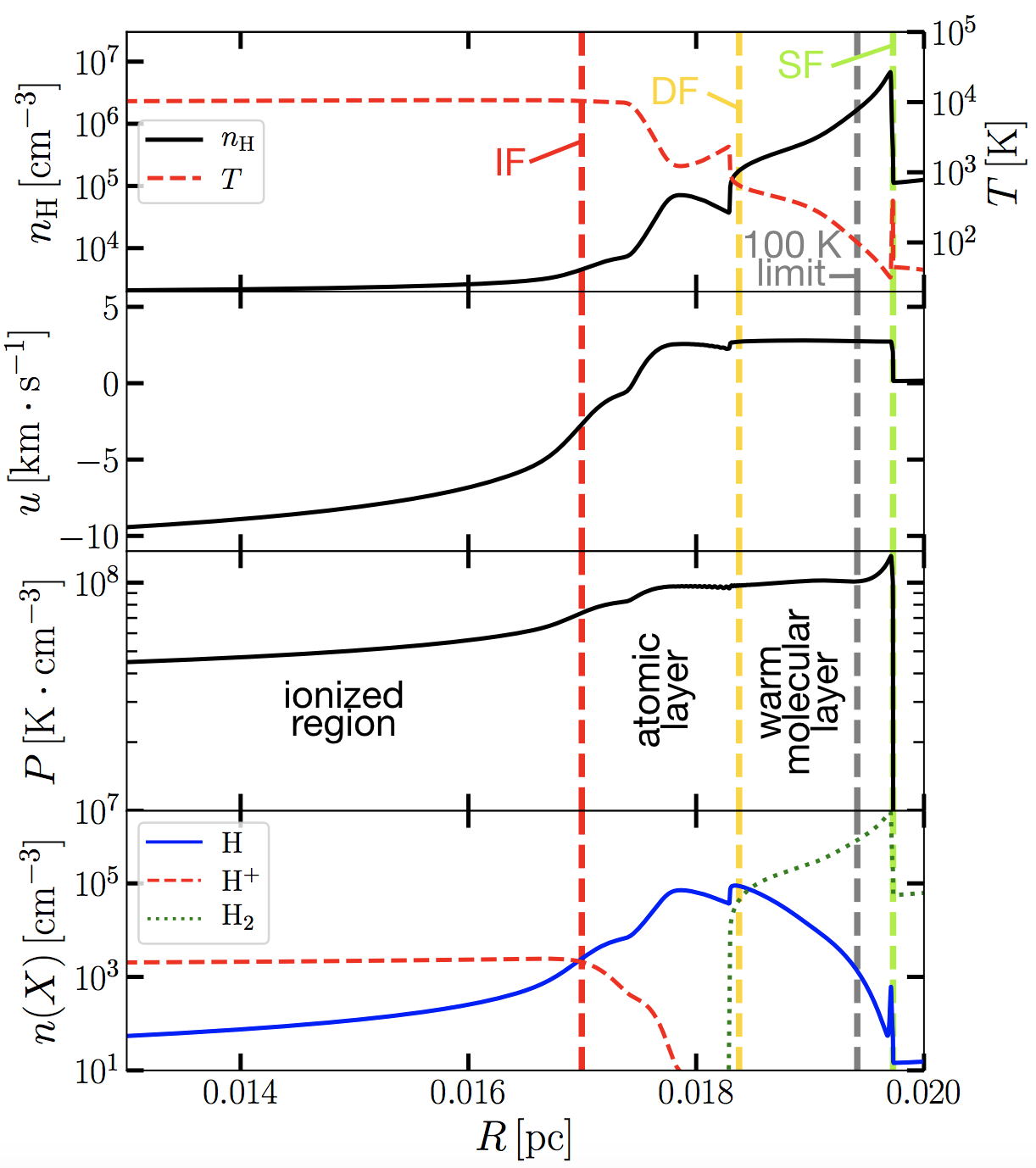}
    \caption{{\bf Left:} Evolution of the structure of the photoevaporating PDR for the example model $n_0=10^5\,\mathrm{cm}^{-3}$, $G_0=10^4$, $T_* = 4\times 10^4 \,\mathrm{K}$. The density and temperature profiles (top panel, black and red line respectively), velocity profile (second panel), pressure profile (third panel), and chemical structure (bottom panel) are shown at five different times (indicated on the first panel). {\bf Right:} Zoom on the profile at the final time (5048 yr). The four panels present the same variables as in the left panels. Also shown are the ionization front (IF, vertical dashed red line), the dissociation front (DF, vertical dashed yellow line), the 100 K limit (vertical dashed grey line) and the shock front (SF, vertical dashed green line). The location of the ionized region, atomic layer, and warm molecular layer (see definitions in text) are shown on the third panel.}
    \label{fig:ExampleRunOne}
  \end{figure*}
}

\newcommand{\FigureExampleRunOneFronts}{%
  \begin{figure}
    \includegraphics[width=1\linewidth]{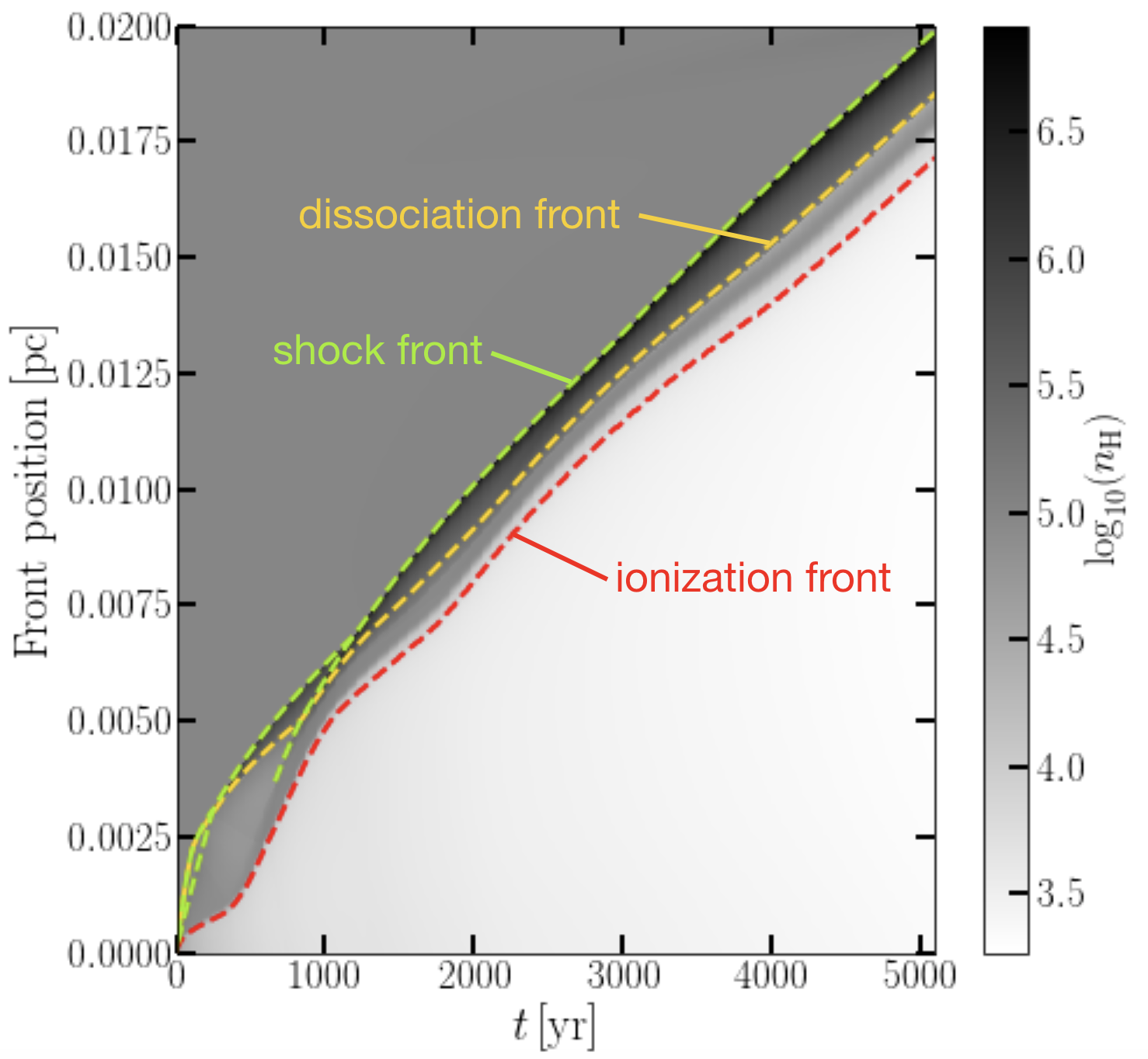}
    \caption{Evolution of the ionization front position (red dashed line), dissociation front position (yellow dashed line), and shock front position (green dashed line) for the example model $n_0=10^5\,\mathrm{cm}^{-3}$, $G_0=10^4$, $T_* = 4\times 10^4 \,\mathrm{K}$, superimposed on the time-position gas density colormap.}
    \label{fig:ExampleRunOneFronts}
  \end{figure}
}

\newcommand{\FigurePGrelation}{%
  \begin{figure}
    \includegraphics[width=1\linewidth]{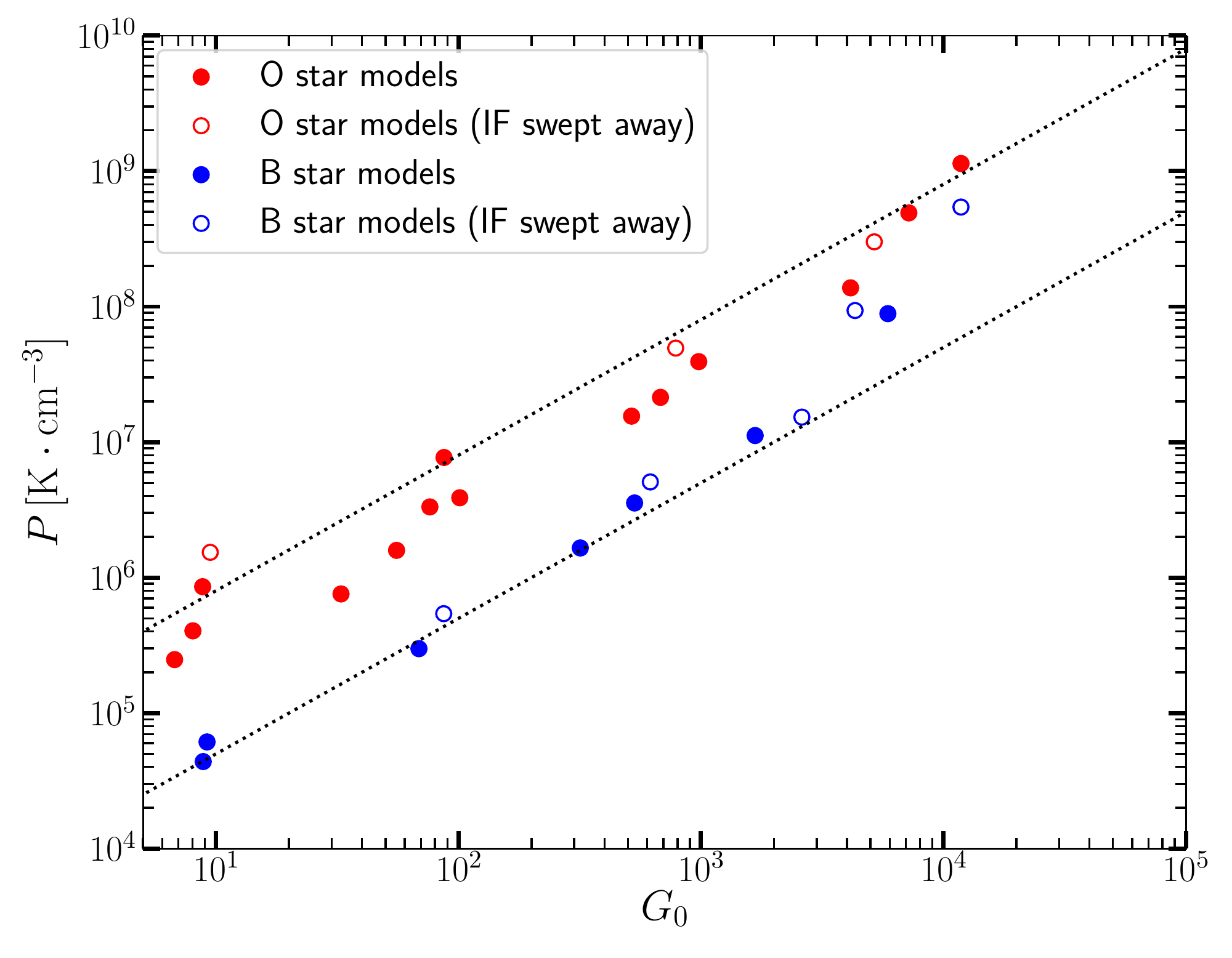}
    \caption{Average thermal pressure in the compressed molecular PDR region versus FUV radiation field at the dissociation front for the final states of our grid of models. O-star models are shown in red and B-star models in blue. Open circles represents models in which the ionization front is swept away by the evaporation flow.}
    \label{fig:PGrelation}
  \end{figure}
}

\newcommand{\FigureScenarioMap}{%
  \begin{figure*}
    \includegraphics[width=0.5\linewidth]{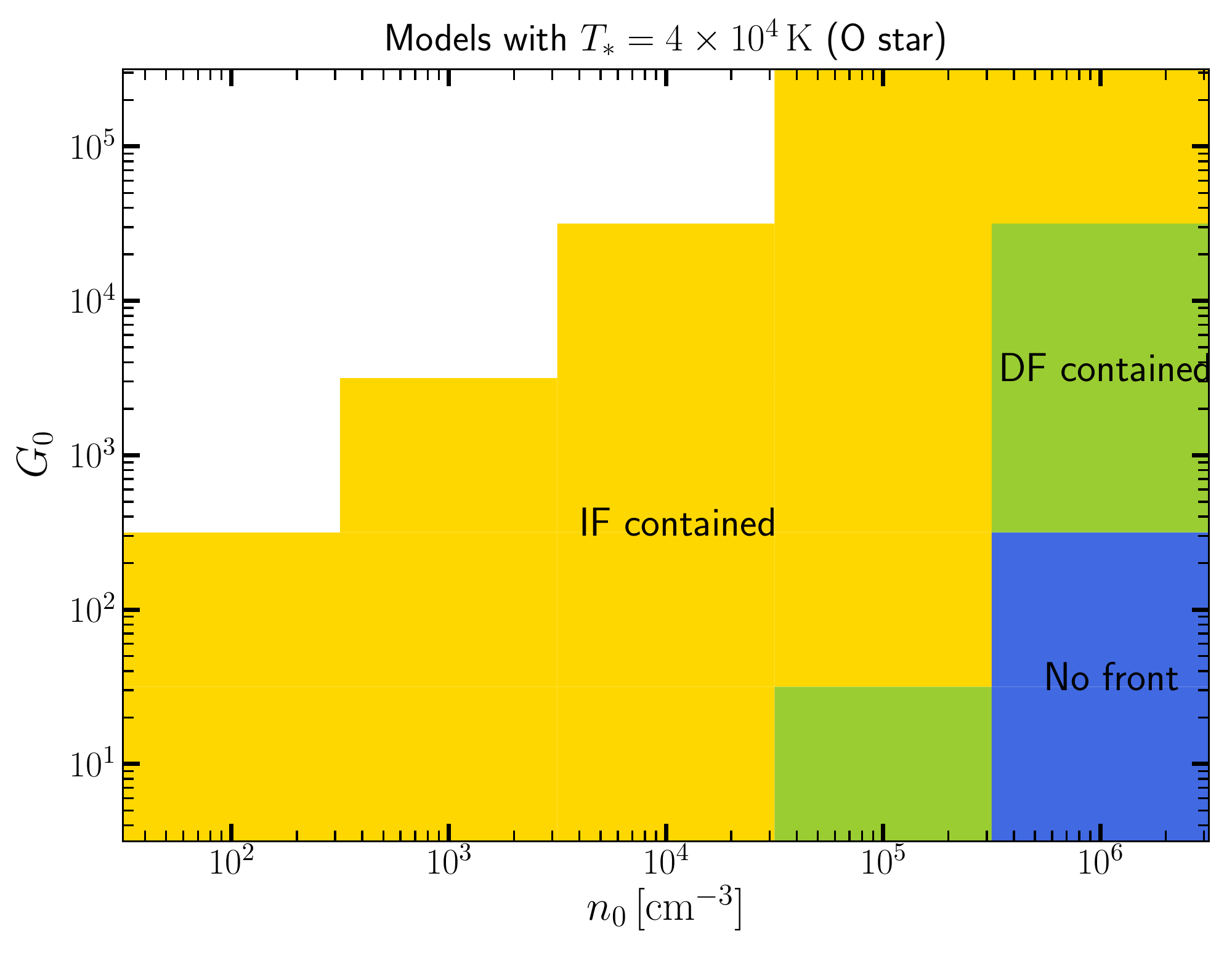}
    \includegraphics[width=0.5\linewidth]{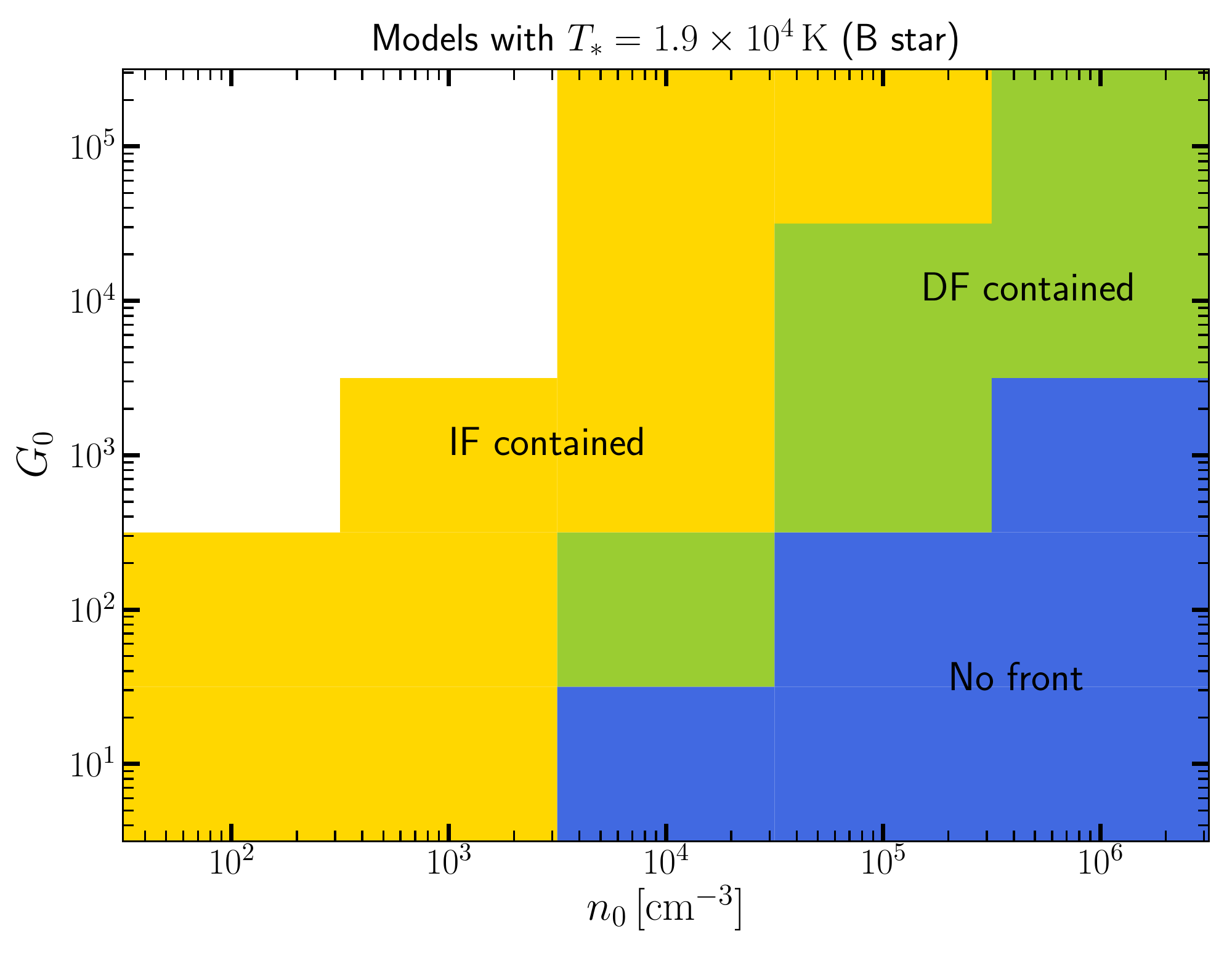}
    \caption{Domains of the different photoevaporation scenario : (1) both the IF and the DF are contained and propagate away from the star (yellow), (2) the IF is swept away towards the star by the evaporation flow but the DF is contained and propagate away from the star (green), (3) both the IF and the DF are swept away along with the evaporation flow (blue), in the $n_0-G_0$ plane for $T_*=4\times 10^4$ K (left panel) and $T_*=1.9\times 10^4$ K (right panel)}
    \label{fig:ScenarioMap}
  \end{figure*}
}

\newcommand{\FigureExampleRunTwo}{%
  \begin{figure}
    \includegraphics[width=1\linewidth]{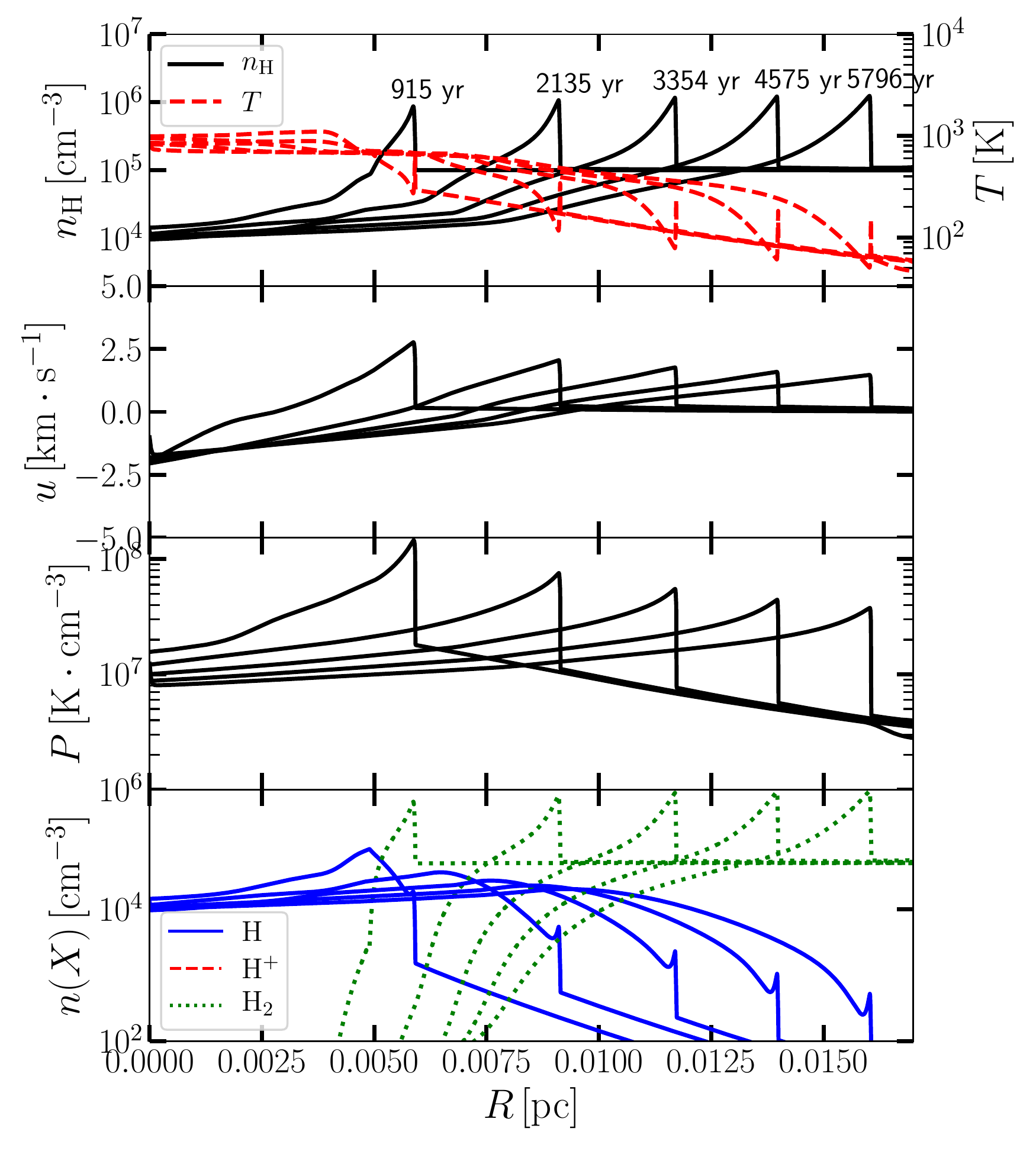}
    \caption{Evolution of the structure of the photoevaporating PDR for the B star example model $n_0=10^5\,\mathrm{cm}^{-3}$, $G_0=10^4$, $T_* = 1.9\times 10^4 \,\mathrm{K}$. The density and temperature profiles (top panel, black and red line respectively), velocity profile (second panel), pressure profile (third panel), and chemical structure (bottom panel) are shown at five different times (indicated on the first panel).}
    \label{fig:ExampleRunTwo}
  \end{figure}
}

\newcommand{\FigureExampleRunTwoFronts}{%
  \begin{figure}
     \includegraphics[width=1\linewidth]{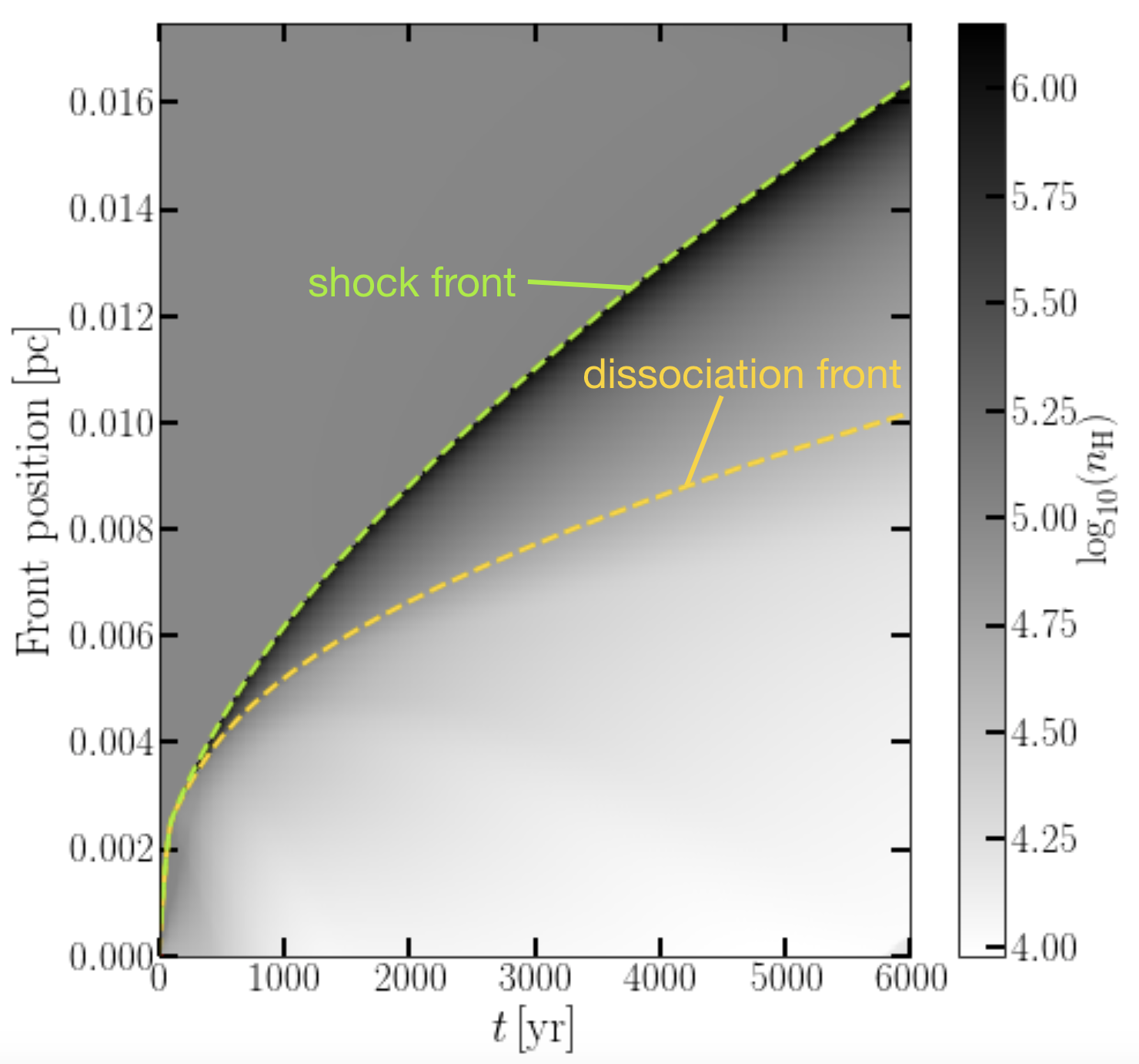}
    \caption{Evolution of the dissociation front position (yellow dashed line), and shock front position (green dashed line) for the B star example model $n_0=10^5\,\mathrm{cm}^{-3}$, $G_0=10^4$, $T_* = 1.9\times 10^4 \,\mathrm{K}$, superimposed on the time-position gas density colormap.}
    \label{fig:ExampleRunTwoFronts}
  \end{figure}
}

\newcommand{\FigurePressureVariations}{%
  \begin{figure*}
     \includegraphics[width=0.5\linewidth]{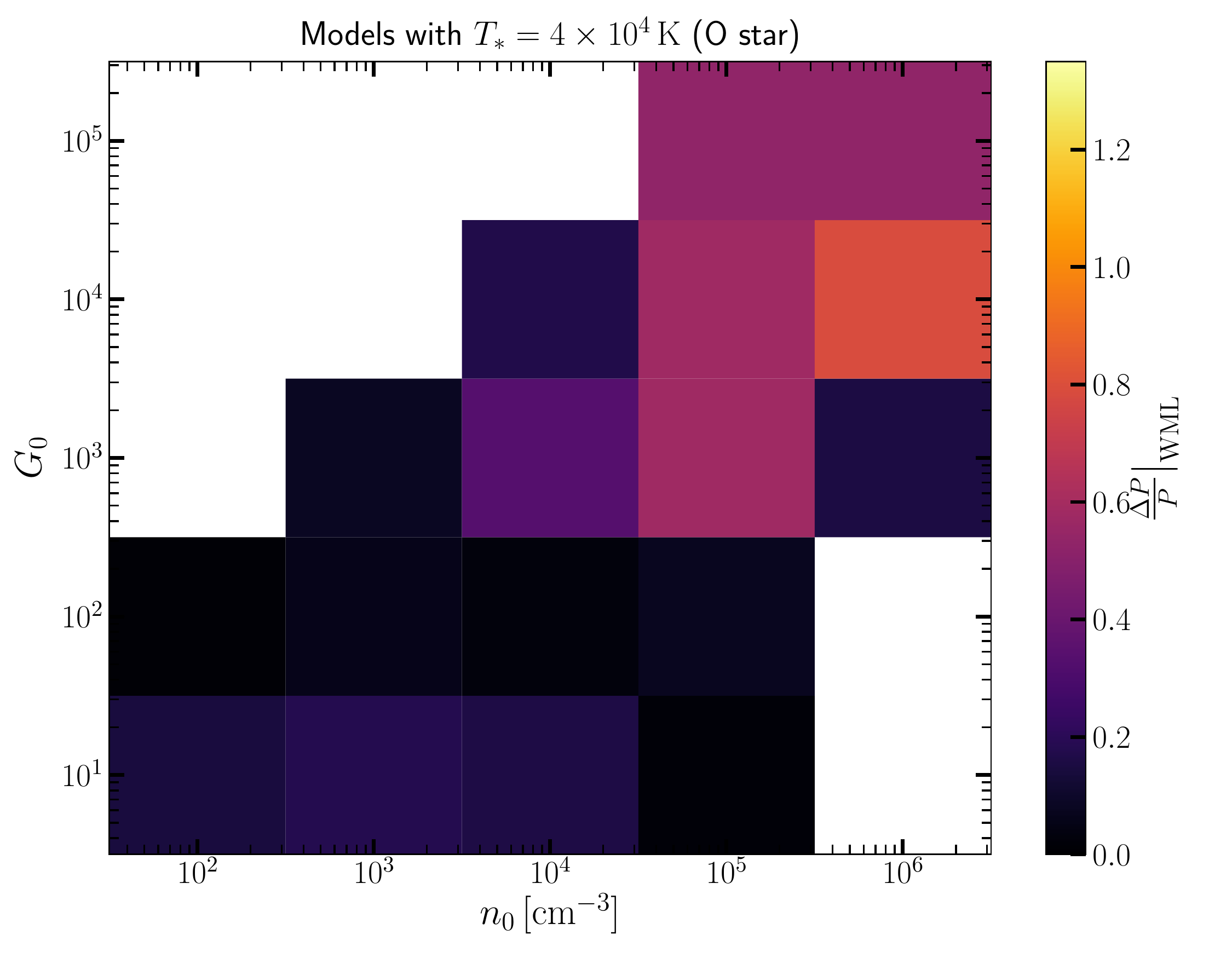}
     \includegraphics[width=0.5\linewidth]{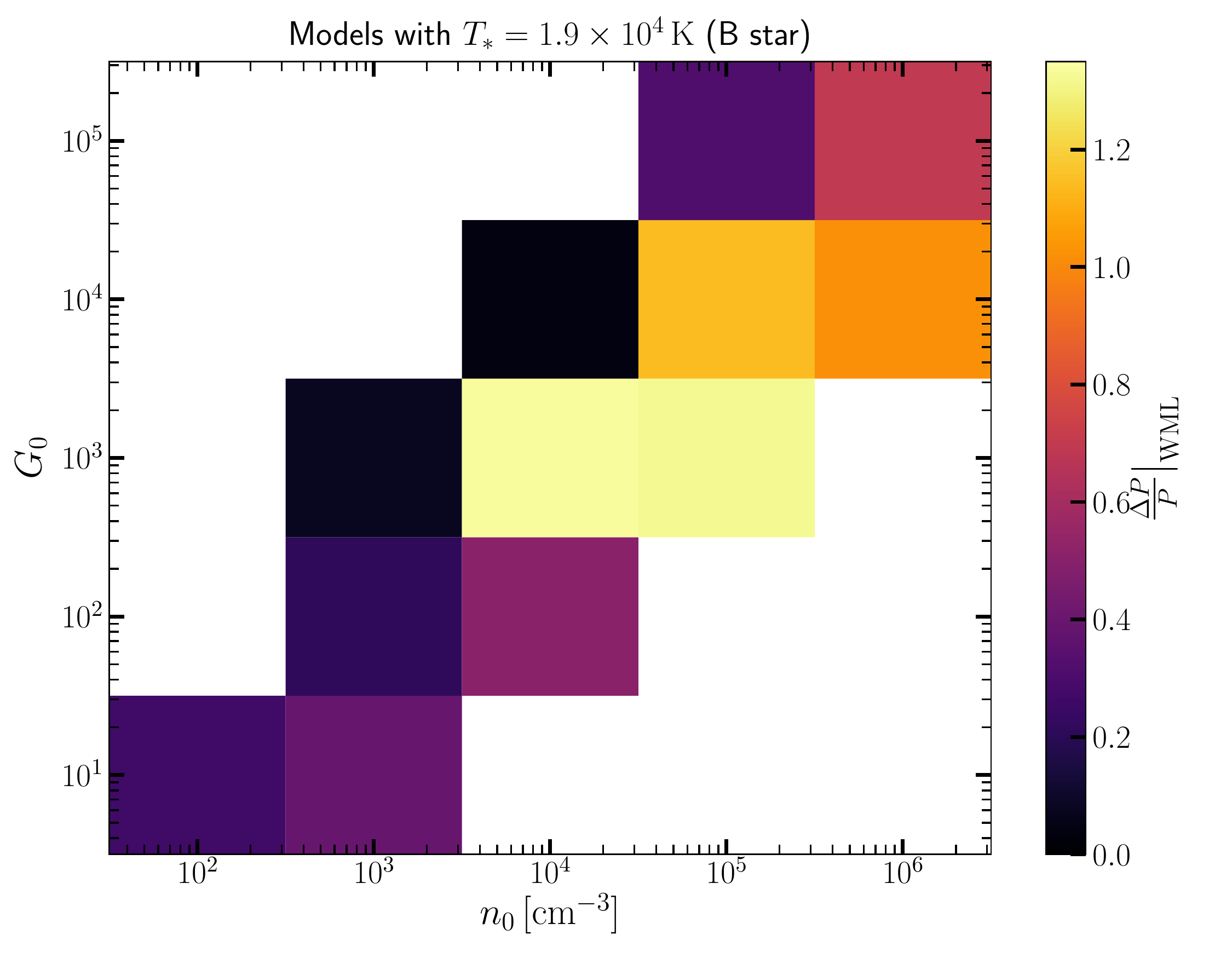}
    \caption{Relative pressure increase $\Delta P / P$ over the warm molecular layer (WML) of the PDR, for the O star models (left) and the B star models (right). }
    \label{fig:PressureVariations}
  \end{figure*}
}

\newcommand{\FigureDensityRatio}{%
  \begin{figure*}
     \includegraphics[width=0.5\linewidth]{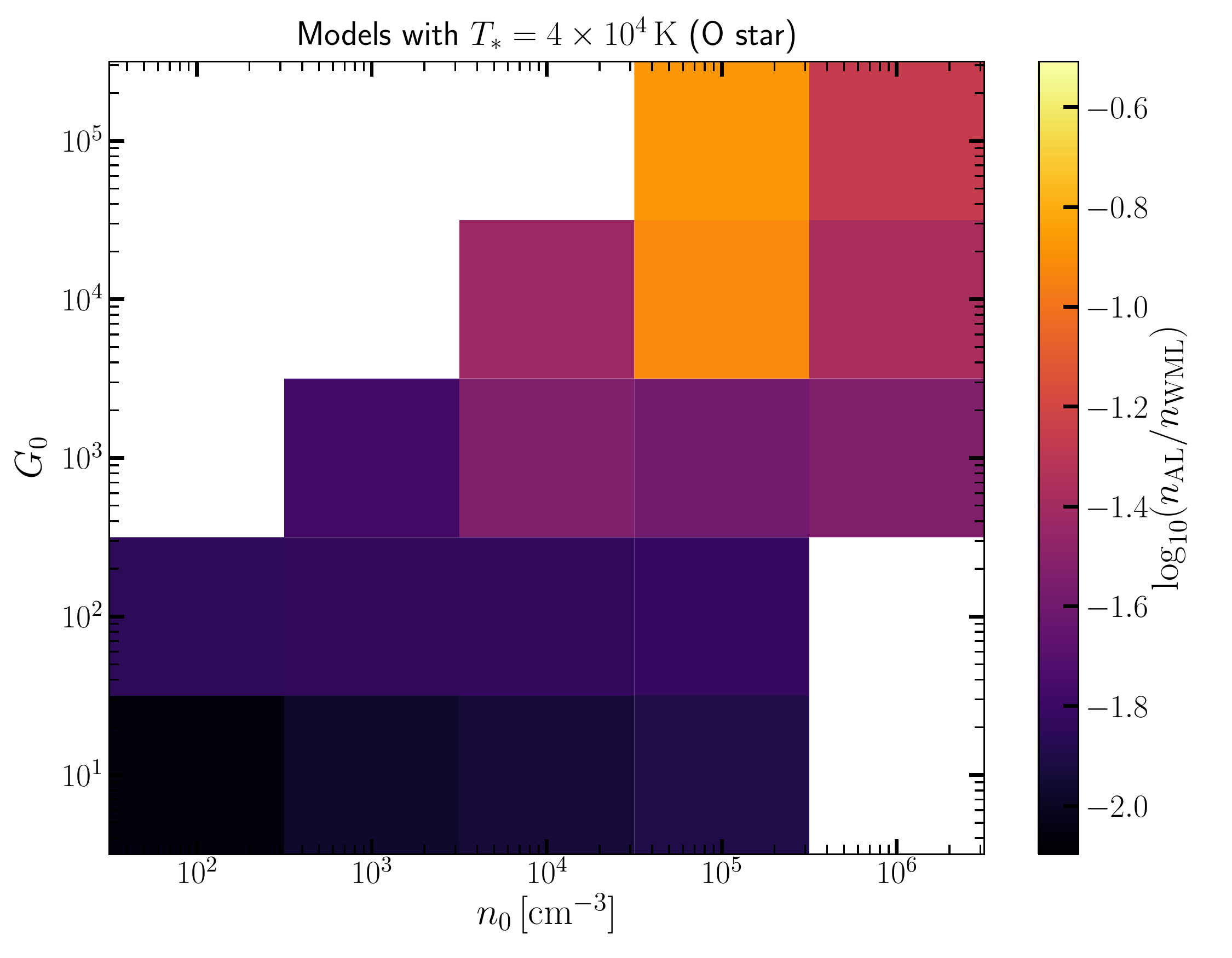}
     \includegraphics[width=0.5\linewidth]{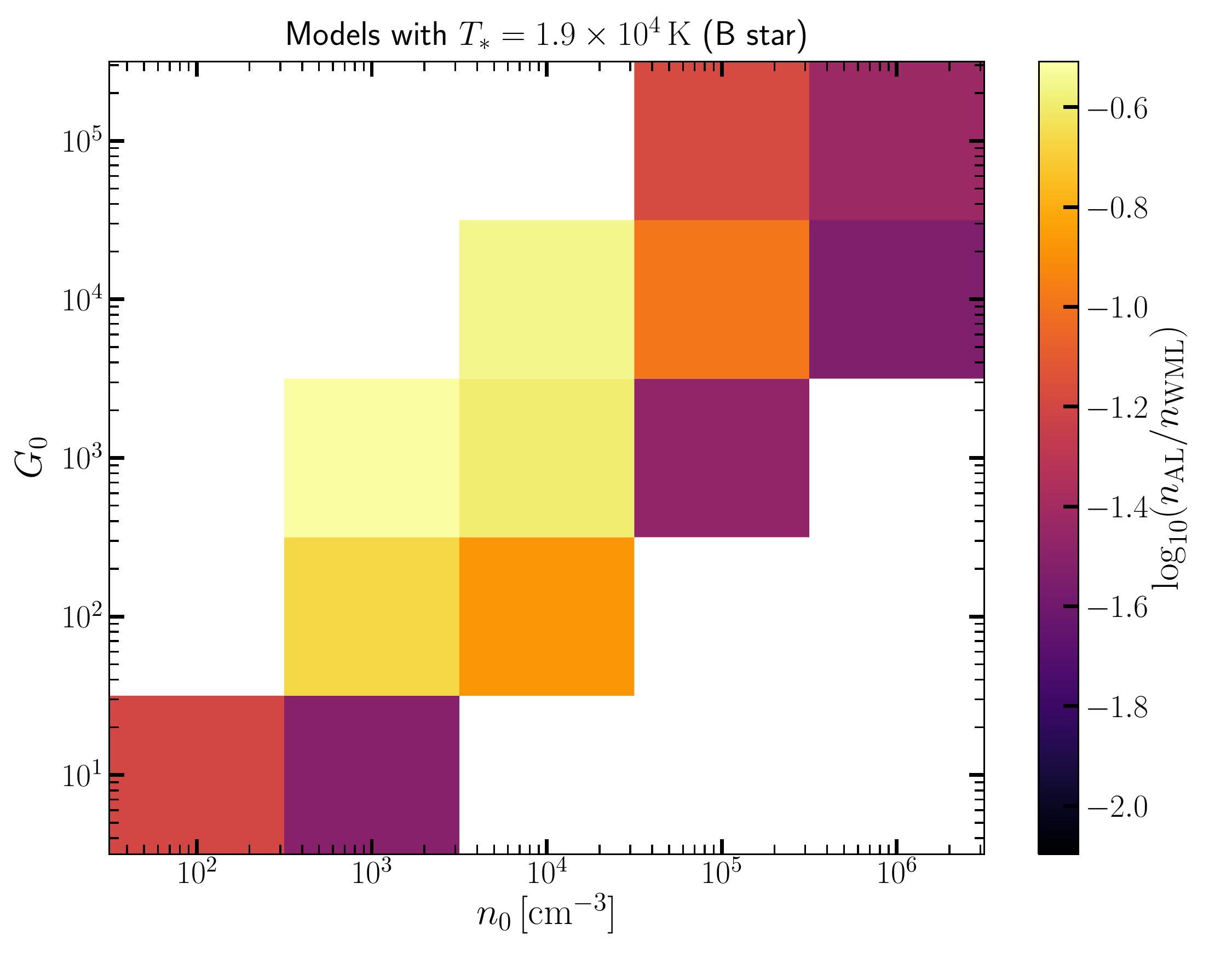}
    \caption{Density ratio between the average gas densities in the atomic layer (AL) and in the warm molecular layer (WML), for the O star models (left) and the B star models (right). }
    \label{fig:DensityRatio}
  \end{figure*}
}

\newcommand{\FigureWMLsize}{%
  \begin{figure*}
     \includegraphics[width=0.5\linewidth]{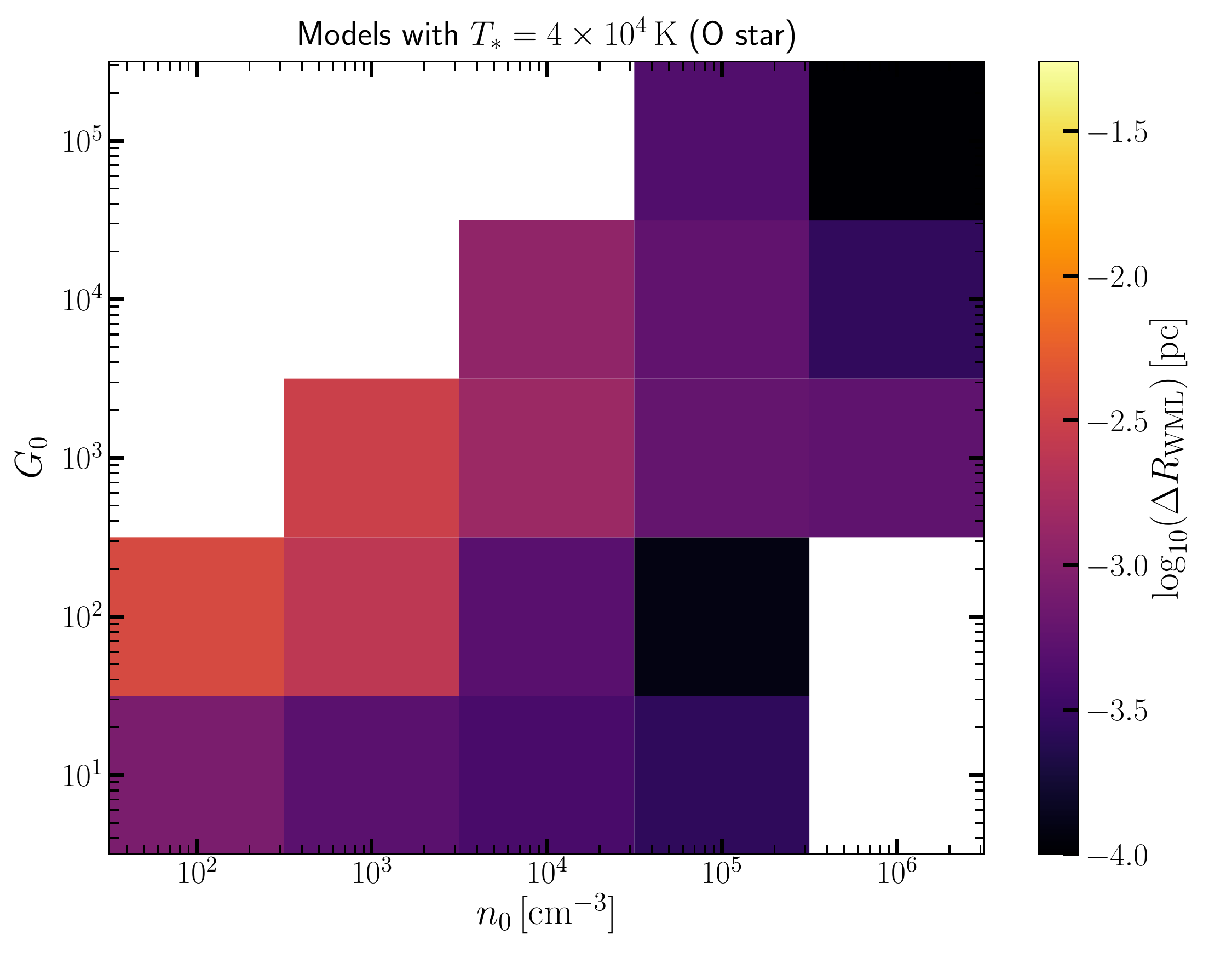}
     \includegraphics[width=0.5\linewidth]{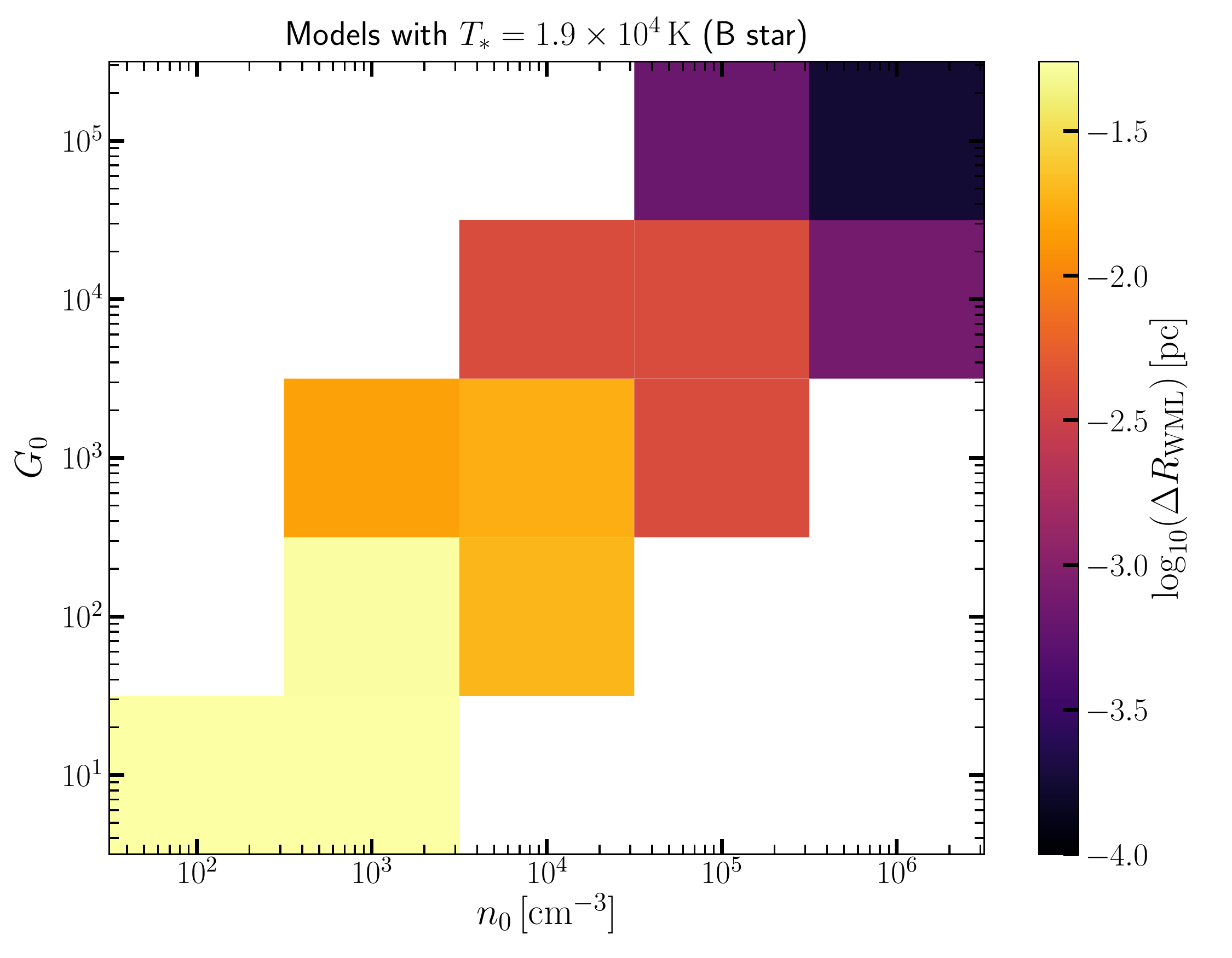}
    \caption{Thickness of the warm molecular layer (WML) of the PDR, for the O star models (left) and the B star models (right). }
    \label{fig:WMLsize}
  \end{figure*}
}

\newcommand{\FigureSizeRatio}{%
  \begin{figure*}
     \includegraphics[width=0.5\linewidth]{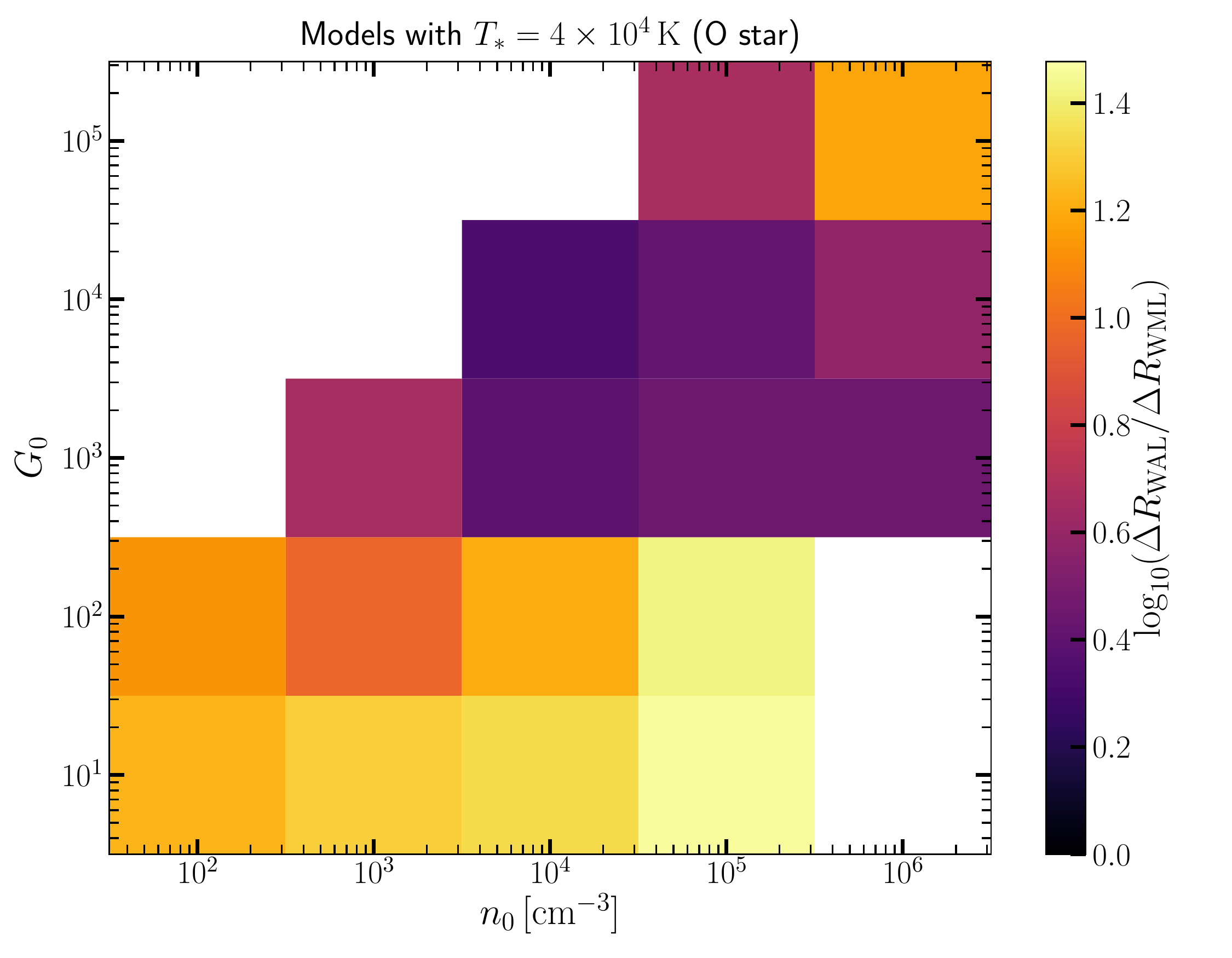}
     \includegraphics[width=0.5\linewidth]{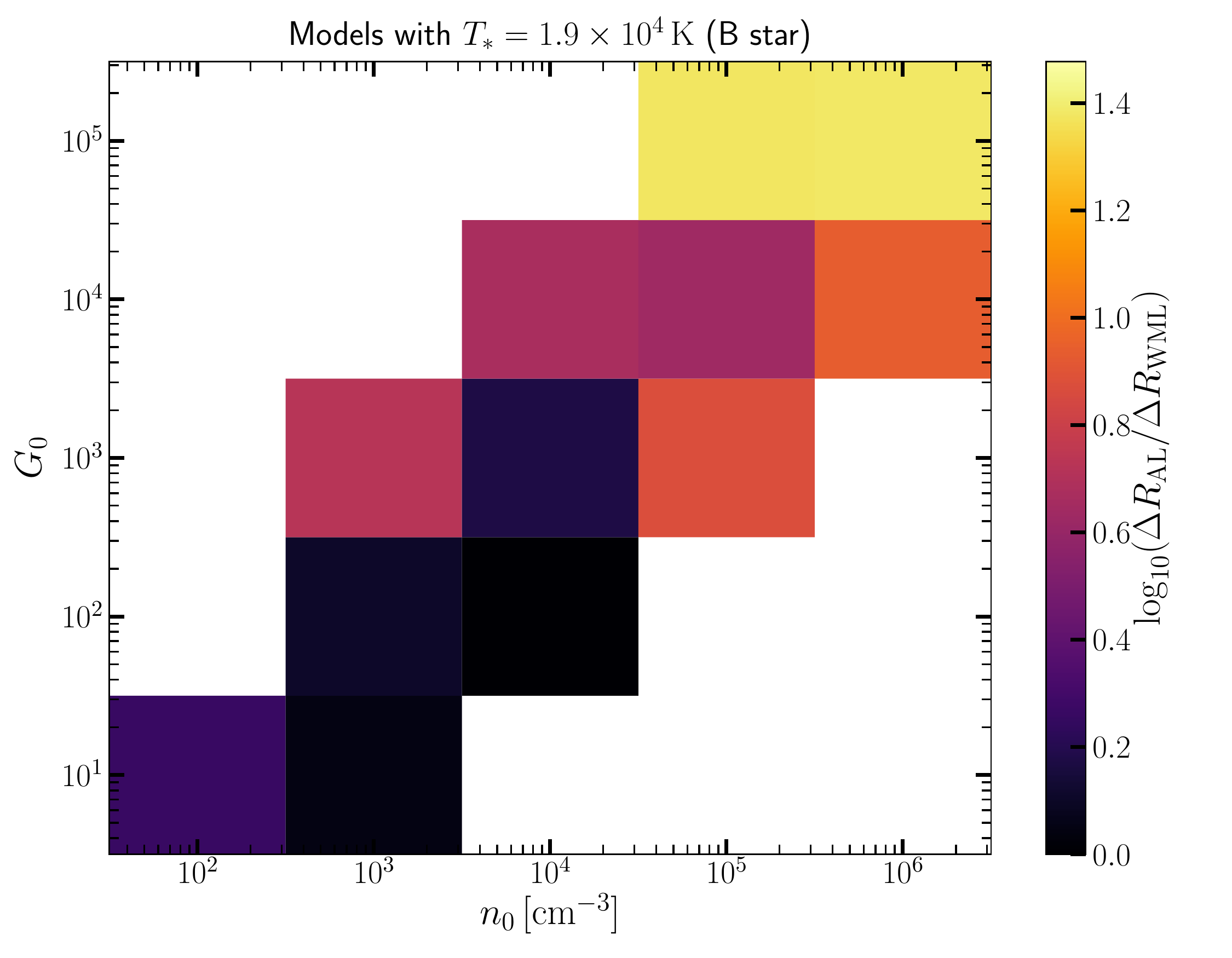}
    \caption{Ratio of the thicknesses of the atomic layer (AL) and of the warm molecular layer (WML) of the PDR, for the O star models (left) and the B star models (right). }
    \label{fig:SizeRatio}
  \end{figure*}
}

\newcommand{\FigureShockVelocities}{%
  \begin{figure*}
     \includegraphics[width=0.5\linewidth]{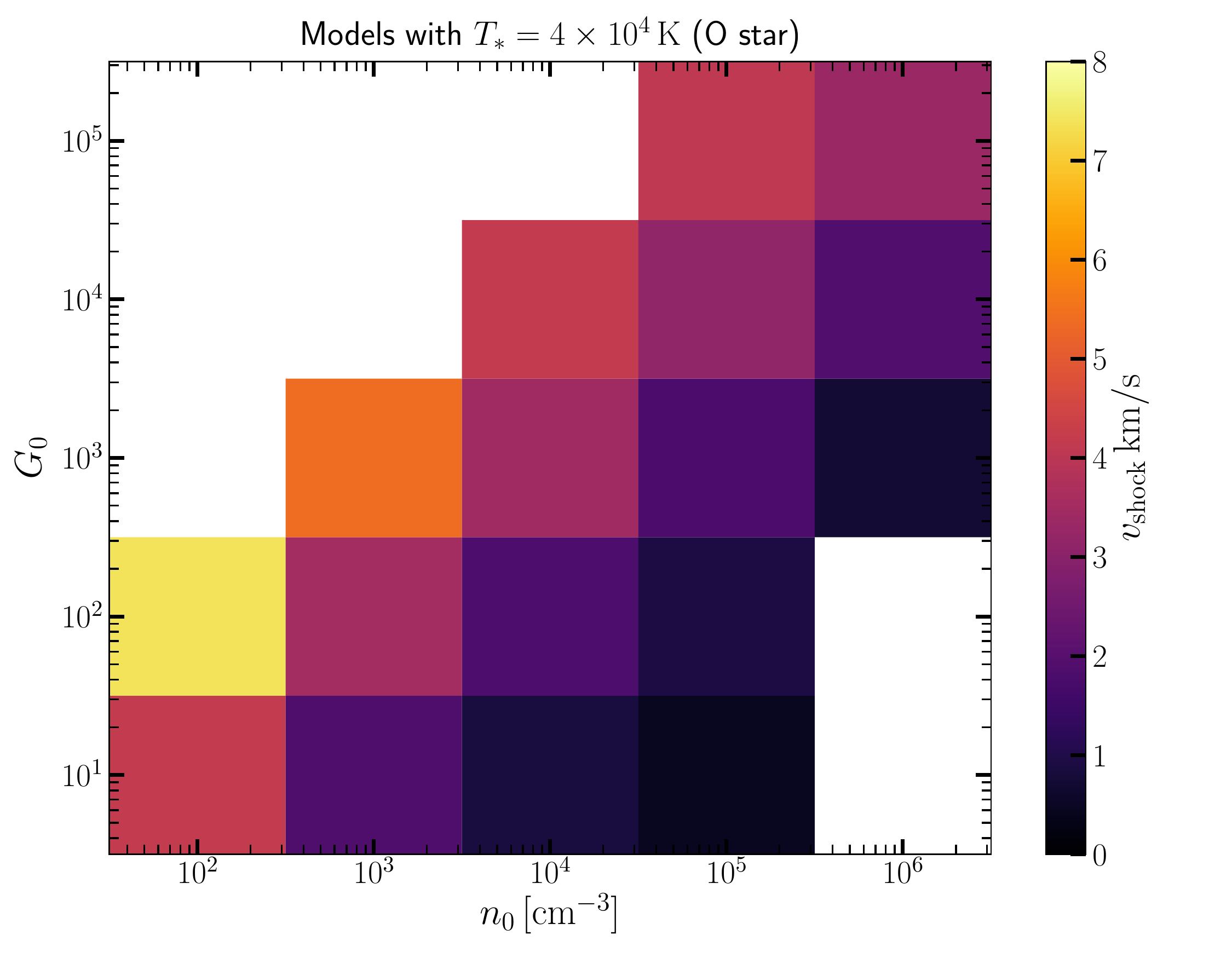}
     \includegraphics[width=0.5\linewidth]{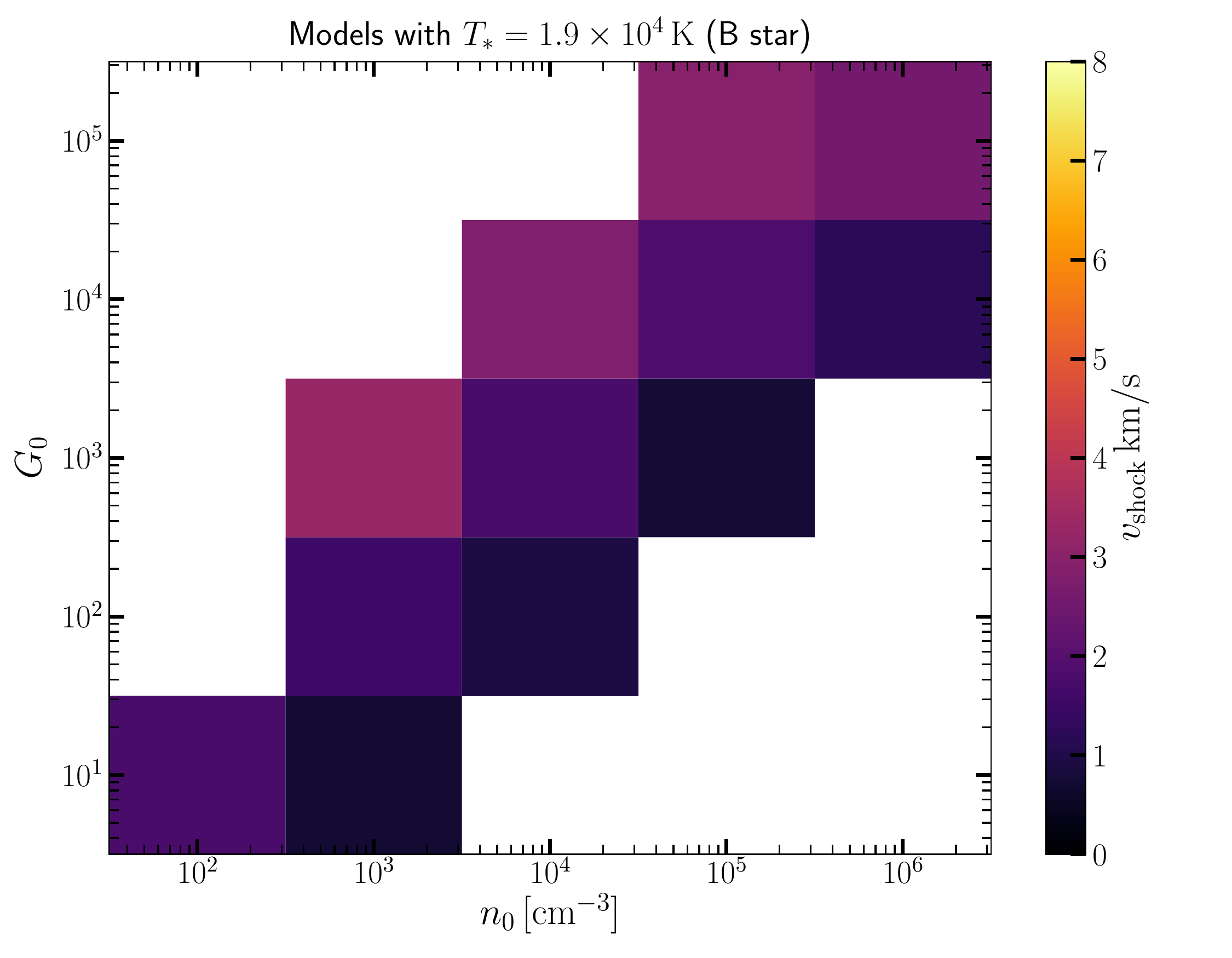}
    \caption{Shock velocity for the O star models (left) and the B star models (right). }
    \label{fig:ShockVelocities}
  \end{figure*}
}

\begin{document}

\title{Photoevaporating PDR models with the \textit{Hydra} PDR Code} %

\author{Emeric Bron\inst{\ref{ICMM}} %
  \and  Marcelino Ag\'undez\inst{\ref{ICMM}} %
  \and Javier R. Goicoechea\inst{\ref{ICMM}} %
  \and Jos\'e Cernicharo\inst{\ref{ICMM}} %
}%

\institute{%
  ICMM, Consejo Superior de Investigaciones Cientificas
  (CSIC). E-28049. Madrid, Spain. \label{ICMM} %
} %

\date{Received ? / Accepted ?} %

\abstract%
{Recent Herschel and ALMA observations of Photodissociation Regions (PDRs) have revealed the presence of a high thermal pressure ($P/k_\mathrm{B}\sim 10^7-10^8\,\mathrm{K\, cm}^{-3}$) thin compressed layer at the PDR surface where warm molecular tracer emission (e.g. CH$^+$, SH$^+$, high-$J$ CO, H$_2$,...) originate. These high pressures (unbalanced by the surrounding environment) and a correlation between pressure and incident FUV field ($G_0$) seem to indicate a dynamical origin with the radiation field playing an important role in driving the dynamics.} %
{We investigate whether photoevaporation of the illuminated edge of a molecular cloud could explain these high pressures and pressure-UV field correlation.} %
{We developed a 1D hydrodynamical PDR code coupling hydrodynamics, EUV and FUV radiative transfer and time-dependent thermo-chemical evolution. We applied it to a 1D plane-parallel photoevaporation
scenario where a UV-illuminated molecular cloud can freely evaporate in a surrounding low-pressure medium. We ran a grid of models exploring a range of initial gas densities, incident radiation fields, and stellar types.} %
{We find that photoevaporation can produce high thermal pressures and the observed $P-G_0$ correlation, almost independently from the initial gas density. In addition, we find that constant-pressure PDR models are a better approximation to the structure of photoevaporating PDRs than constant-density PDR models, although moderate pressure gradients are present. Strong density gradients from the molecular to the neutral atomic region are found, which naturally explain the large density contrasts (1-2 orders of magnitude) derived from observations of different tracers. The photoevaporating PDR is preceded by a low velocity shock (a few km/s) propagating into the molecular cloud.} %
{Photoevaporating PDR models offer a promising explanation to the recent observational evidence of dynamical effects in PDRs. Further developments of the model complemented by high angular resolution ALMA and JWST observations will be necessary to constrain the details of the photoevaporation mechanism and to quantitatively understand its impact on observable PDR tracers and more globally on cloud evolution.}

\maketitle %

\section{Introduction}

Photodissociation regions (PDRs, \citealt{Hollenbach1997,Hollenbach1999} for a review) are interface layers where neutral atomic or molecular clouds are exposed to stellar UV illumination. 
PDRs at the surface of strongly irradiated dense molecular clouds are of particular interest because of the rich line and continuum spectra they emit. 
Such dense PDRs are found in star forming regions, resulting from the radiative feedback of newly born massive stars on their parent cloud, and their
emission dominates the IR spectrum of galaxies. They are also found in (proto-)planetary nebulae (e.g. \citealt{Cernicharo2004}), at the opposite end of the stellar lifecycle. A correct understanding of the mechanisms inducing and controlling their emission is thus crucial to probe the star formation cycle of other galaxies. In addition, their rich emission spectra providing a wealth of observational constraints make them ideal laboratories to study the key  processes that govern the physics and chemistry of the interstellar medium (ISM) throughout its evolution, from diffuse interstellar clouds to protoplanetary disks.

Atomic and molecular emission from PDRs has been studied with relative success using stationary PDR models for a long time \citep{Tielens1985,Sternberg1989,Kaufman1999,Meijerink2005,LePetit2006,Rollig2013}. Comparison of the emission from different tracers and lines lead to recognize early the presence of strong density and temperature contrasts in PDRs (e.g., \citealt{Stutzki1988}). This lead to a clumpiness hypothesis for the structure of PDRs, in which high density clumps are embedded in a low density interclump medium (\citealt{Burton1990}). A $\sim0.1\,\mathrm{pc}$-scale clumpiness was initially considered (e.g., \citealt{Stutzki1990}), similar to the more general clumpiness of molecular clouds. But the comparison of the observed spatial size of excited line emission to the predicted PDR thickness in constant density PDR models (at the high densities required by the most excited lines) lead to further hypothesize the existence of sub-mpc scale clumps with several orders of magnitude density contrast compared to the interclump medium (\citealt{Parmar1991}). While this hypothesis has further been used to interpret PDR observations with success (\citealt{Kramer2008}), it was alternately shown that the presence of a density gradient in the PDR, such as induced by constant pressure rather than constant density PDR models, could naturally explain the observations without need for a small-scale clumpiness hypothesis (\citealt{Marconi1998,Habart2005,Allers2005,Nagy2013,Nagy2017}). This isobaric PDR hypothesis has in particular been used by \cite{Joblininprep} to simultaneously explain several tens of atomic and molecular lines in PDRs NGC~7023~NW and the Orion Bar.

These studies have however been mostly based on low spatial resolution observations (>10"), except for ground-based infrared vibrational H$_2$ observations. Recently, ALMA observations \citep{Goicoechea2016,Goicoechea2017} have revealed the spatial structure of the Orion Bar PDR to an unprecedented level of detail. These observations reveal the existence of a thin hot compressed layer of a few arcseconds where hot molecular tracers (reactive ions such as SH$^+$ and HOC$^+$, vibrational H$_2$) are emitted. The authors suggest a dynamical origin to this compressed layer.

Independently of the scenario (clumps, isobaric interface), high thermal pressures (a few $10^7-10^8\,\mathrm{K\, cm}^{-3}$) have been derived from observations of excited lines (especially from Herschel observations) in strongly UV-illuminated galactic PDRs \citep{Allers2005,Perez-Beaupuits2010,Goicoechea2011,Goicoechea2016,Nagy2012,Nagy2013,Nagy2017,Joblininprep}. While the high pressures inside compact H\textsc{ii} regions (which are confined by the surrounding molecular cloud) can induce compression of the surrounding PDR \citep{Hill1978,Hosokawa2006b}, the most studied galactic PDRs either surround H\textsc{ii} regions that have already broken out of their parental molecular clouds (at least partly, e.g. the Orion Bar) or surround B stars that only have very small H\textsc{ii} regions (e.g. NGC~7023). Compression of the PDR by the pressure-driven expansion of the H\textsc{ii} region thus cannot explain the high PDR pressures in these objects. The pressure in the central ionized or neutral cavities that these PDRs surround is found to be significantly lower than in the hot molecular layer (e.g. $P/k_\mathrm{B}=6\times 10^7\,\mathrm{K\, cm}^{-3}$ vs. $2 \times 10^8\,\mathrm{K\, cm}^{-3}$ in the Orion Bar, \citealt{Goicoechea2016}), indicating that these high pressures could only be sustained dynamically.

When the surface layer of photo-heated gas is free to expand towards the star, this expansion exerts a force on the molecular cloud that can
induce compression of an interface layer at the edge of the molecular cloud. This process is called photoevaporation and was investigated as a clump acceleration mechanism as early as \cite{Oort1955}. A detailed analytical theory of photoevaporation fronts has been developed by \cite{Bertoldi1989,Bertoldi1996}. While photoevaporation has mainly been studied in the context of isolated cometary globules inside H\textsc{ii} regions (e.g. \citealt{Lefloch1994,Lefloch1995}),  
photoevaporative compression could play a central role in controlling the density and pressure structure of PDRs at the irradiated edges of larger molecular clouds, and hold the key to explaining the density contrasts deduced from the observations of different PDR tracers. 
In addition, recent Herschel observations seem to show a strong correlation between pressure in the PDR and impinging FUV radiation field \citep{Chevance2016,Joblininprep}, providing further evidence for photoevaporation being the control mechanism of the pressure structure of PDRs.

The current state-of-the-art PDR models are stationary (stationary chemistry and no consideration of gas dynamics) and cannot properly include photoevaporative compression nor the resulting non-stationary chemical effects \citep{Storzer1998}. In contrast, numerous 3D hydrodynamical/MHD codes including some level of photoionizing radiation transfer have been developed (e.g. \citealt{Henney2005,Arthur2011,Mackey2011,Haworth2012,Rosdahl2013,Tremblin2014,Baczynski2015}) but include no (or very limited) PDR physics and chemistry, as they focus mainly on the dynamics of the ionization front itself or on the larger scale dynamics of star formation.
We aim at bridging this gap by expanding PDR models to include the minimal additional physics to properly describe photoevaporation of PDR fronts, while retaining the capability to treat detailed PDR microphysics and chemistry with a limited computation time allowing to explore large parameter spaces with grids of models. We developed a hydrodynamical PDR model coupling 1D hydrodynamics, time-dependent thermal and chemical evolution, and UV radiative transfer, which we use to study the pressure and density structure of photoevaporating PDRs, accounting for photoevaporation induced by both photoionization heating and by PDR heating (photoelectric effect and  collisional de-excitation of UV-pumped H$_2$). We focus here on a photoevaporation scenario in which the edge of a dense molecular cloud is exposed to stellar UV photons and can freely evaporate into a low pressure cavity, corresponding to PDRs surrounding late phase H\textsc{ii} regions, after the H\textsc{ii} bubble has broken out of its parental molecular cloud.
 A few similar codes have been previously published \citep{Hosokawa2006b,Kirsanova2009}, but only applied to pressure-driven expanding H\textsc{ii} regions and their surrounding PDRs.
\cite{Motoyama2015} developed a 2D hydrodynamical PDR code, but do not include hydrogen photoionization. \cite{Gorti2002} have also investigated the photoevaporation of small spherical clumps.

In Sect. \ref{sec:Code}, we present the new \textit{Hydra} PDR code, which we validate on test problems in Sect. \ref{sec:Tests}. Our results on the structure of photoevaporating PDRs are presented in Sect. \ref{sec:Results}. We then discuss the limitations of our model in Sect. \ref{sec:Discussion} before presenting our conclusions in Sect. \ref{sec:Conclusion}.

\section{A dynamical PDR code}\label{sec:Code}

The \textit{Hydra} PDR Code couples compressible hydrodynamics, chemical and thermal evolution of the gas, 
and radiative transfer in 1D geometries (planar or spherically symmetric). Both the hydrodynamical
variables of the gas: the gas velocity $u$, the mass density $\rho$ and the total (mechanical plus internal)
energy per unit mass $E$; and the chemical mass fractions, noted $y_i$ for species $i$, are followed as a
function of time and position
during the evolution of the photoevaporating ionization/PDR front.

We use Lagrangian coordinates, and use the mass coordinate $dm = \rho r^\alpha dr$ where 
$\alpha = 0$ in a planar geometry, $\alpha=1$ in cylindrically-symmetric geometry and  $\alpha = 2$ in a spherically-symmetric geometry.
The evolution equations are given by mass conservation, momentum conservation, total energy (mechanical plus internal) conservation, and chemical
mass fraction conservation:
\begin{eqnarray}
\frac{d}{dt}\left(\frac{1}{\rho}\right) - \frac{\partial \left(r^\alpha u \right)}{\partial m} & = & 0 \label{eq:mass_conservation}\\
\frac{du}{dt} + r^\alpha \frac{\partial P}{\partial m} & = & 0 \label{eq:momentum_conservation}\\ 
\frac{dE}{dt} + \frac{\partial \left( r^\alpha\,u\,P \right)}{\partial m} & = & \frac{1}{\rho}\left( \Gamma - \Lambda \right) \label{eq:energy_conservation}\\
\frac{dy_i}{dt} & = & \frac{m_i}{\rho}\left(F_i - D_i\right) \label{eq:mass_fractions_conservation}
\end{eqnarray}
where $P$ is the thermal pressure, $\Gamma$ and $\Lambda$ are the total radiative and chemical heating 
and cooling rates per unit volume, $m_i$ is the mass of species $i$ and $F_i$ and $D_i$ are
the formation and destruction rates (in number per unit volume) of species $i$.
The pressure $P$ is related to the primary variables by the equation
\begin{equation}
P = (\gamma - 1)\rho(E-\frac{u^2}{2})
\end{equation}
where $\gamma$ is the adiabatic index (5/3 for a monoatomic gas, 7/5 for a diatomic gas).

In addition, the heating/cooling rates and formation/destruction rates are
more easily expressed as functions of the species number densities $n_i$ and
the gas temperature $T$. These variables are expressed as a function of
the primary variables as
\begin{eqnarray}
n_i & = & \frac{\rho}{m_i} y_i \\
T & = & \frac{P}{k_B\sum_i n_i}
\end{eqnarray}
Radiation pressure and gravitational forces are not included in this paper. Indeed,
\cite{Krumholz2009} have shown radiation pressure to have negligible impact except around 
large massive star clusters, and gravitational forces are negligible for the purpose
of studying the structure of localized ionization/dissociation fronts.

\FigureCodeStructure

We use simple operator splitting (Godunov splitting, see e.g. Chap. 17 in \citealt{leveque_2002}) to solve separately the hydrodynamical evolution (corresponding 
to Eq. \ref{eq:mass_conservation}, \ref{eq:momentum_conservation},
\ref{eq:energy_conservation} and \ref{eq:mass_fractions_conservation} without the right-hand sides) and the thermochemical
evolution of the gas (evolution of the thermal and chemical state of the gas defined by the right-hand sides of the same equations). 
The hydrodynamical evolution of the variables is thus first computed over one full time step duration, after which the thermochemical
evolution is computed over the same duration, but starting from the final state of the hydrodynamical step.
As the thermochemical evolution depends on non-local radiative effects, the thermochemical step is preceded
by a radiative computation of the necessary quantities. The different stages of one global time step are 
summarized on Fig. \ref{fig:CodeStructure}, and are discussed in more details in the subsections below. 

We chose to restrict ourselves to 1D geometries in order to keep a reasonable computation time allowing 
us to run grids of models exploring large parameter spaces. 
The possible geometries are either planar (e.g. a plane-parallel ionization/photodissociation front) 
or spherically symmetric (e.g. the pressure-driven expansion of a spherical H\textsc{ii} region, 
or an isotropically illuminated spherical cloud).

\subsection{Hydrodynamical solver}

The hydrodynamical stage of the time step is computed by solving the ideal compressible hydrodynamics equations
(Eq. \ref{eq:mass_conservation}, \ref{eq:momentum_conservation} and \ref{eq:energy_conservation}, without right hand side).
We use the Piecewise Parabolic Method (PPM) of \cite{Colella1984}, a higher order Godunov-type method 
of the MUSCL family \citep{VanLeer1979}.

The simulation domain is discretized on a Lagrangian grid, whose cells move and contract/expand along with the gas. 
This allows for the spatial resolution to automatically increase in the compressed layer immediately behind the ionization/dissociation 
front. This is crucial as the thickness of the compressed PDR layer is often found to be very small compared to the distance
travelled by the ionization/dissociation front during its development. In addition, this Lagrangian scheme also avoids any numerical 
diffusion of the chemical abundances across sharp photochemical transitions, as the chemical mass fractions are exactly 
conserved during the hydrodynamical step.

\subsection{Thermal and chemical evolution}

The thermochemical stage of the global time step computes the evolution induced by the right-hand sides of
Eq. \ref{eq:energy_conservation} and \ref{eq:mass_fractions_conservation}. Only $y_i$ and $E$ vary
during this evolution. For simplicity, we actually write the evolution equations for this stage in terms
of the number densities $n_i$ and of the thermal energy (volumetric) density $e = \rho(E - u^2/2)$:
\begin{eqnarray}
\frac{de}{dt} & = & \Gamma - \Lambda \\
\frac{dn_i}{dt} & = & F_i - D_i
\end{eqnarray}

This system of ODEs is integrated in each cell for the duration of the time step using the stiff ODE solver 
CVODE \citep{Hindmarsh2005}, which uses Backwards Differentiation Formulas with automatically varying 
order, and automatically subdivides the time step into substeps. Elemental abundance conservation is not
enforced during the chemical time step, but is verified and corrected at the end of the time step. The deviation from
elemental conservation during one chemical time step (before correction) is found to be at most 
a relative error of $\sim 10^{-9}$ (relative to the total abundance of the element, the largest errors being usually found for O or C conservation).

\paragraph{Chemical network -}
In this paper, the gas phase chemical network is limited to a reduced PDR and ionization front chemistry, allowing
to follow the ionization front, the H/H$_2$ transition and the C$^+$/C/CO transition. The 28 species included are
H, H$_2$, C, O, He, Fe, Na, Mg, CH, OH, H$_2$O, CO, e$^-$, H$^+$, C$^+$, O$^+$, H$_2^+$,
H$_3^+$, Fe$^+$, Na$^+$, Mg$^+$, CH$^+$, OH$^+$, H$_2$O$^+$, H$_3$O$^+$, CO$^+$ and 
HCO$^+$. The chemical network includes 189 chemical reactions and is the subset of the chemical
network used by \cite{Agundez2017} induced by our limited species list, with the following
modifications. The photodissociation of H$_2$ and CO are replaced by the treatment described in
the following subsection. The only included surface reaction on dust grains is the formation of H$_2$ on dust grains
using the simple rate expression 
\begin{equation}
R_{\mathrm{H}_2} = 3\times 10^{-17} n_\mathrm{H}n(\mathrm{H}) \sqrt{\frac{T}{70\,\mathrm{K}}} \frac{s_\mathrm{H}(T)}{s_\mathrm{H}(70\,\mathrm{K})}\,\mathrm{cm^{-3}\, s^{-1}}
\end{equation}
where $s_\mathrm{H}(T)$ is the sticking coefficient for hydrogen atoms used in \cite{LeBourlot2012}:
\begin{equation}
s_\mathrm{H}(T) = \frac{1}{1 + \left( \frac{T}{464\,\mathrm{K}}\right)^{\frac{3}{2}}}
\end{equation}
which gives results similar to the expression of \cite{Sternberg1995}.
The treatment of the EUV photoionization of H is described in the next subsection.
For hydrogen recombination, we assume that all recombinations to the fundamental
level result in another photoionization on the spot, so that we take for the effective recombination
rate the "case B" recombination coefficient including only recombinations to levels $n\ge 2$ (cf. 4.2.a in \citealt{Spitzer1978}), for 
which we use the expression from Table 7.3 of \cite{Tielens2005}. 
Radiative and dielectronic recombination coefficients for C$^{+}$ and O$^{+}$ are also taken 
from \cite{Tielens2005}.
Except for H$_2$ and CO photodissociation and the EUV photoionization reaction of H (discussed in the next subsection), 
all photo-reactions use $A_\mathrm{V}$-dependent expressions, where $A_\mathrm{V}$ (counted from the star to the current position) is estimated
during the radiative transfer stage.

\paragraph{Heating processes -}
In the ionized region, the heating is dominated by H photoionization. In the neutral compressed PDR, the heating 
mechanisms included are the photoelectric 
effect (following the prescription of \citealt{Bakes1994}), heating by H$_2$ UV pumping (using the approximate 
expression of \citealt{Rollig2006}), heating by H$_2$ formation on grains (using the formalism from \citealt{Hollenbach1979}, 
but updating the repartition of the formation energy to the values of \citealt{Sizun2010}, assuming that Eley-Rideal formation 
dominates in PDRs), and heating by H$_2$ photodissociation (assuming 0.4 eV by dissociation, \citealt{Hollenbach1979}). Cosmic ray heating is included 
\citep{Tielens2005} to describe the temperature of the molecular gas before being perturbed by the 
ionization/dissociation/shock front.

\paragraph{Cooling processes -}
The cooling inside the ionized region is usually dominated by O\textsc{iii} line cooling, for which we use the expression from
\cite{Tielens2005} (which includes both the lines at 4959 and 5007 \AA{} in the visible, and the FIR fine structure lines at 52 and 88 $\mu$m) assuming
that all oxygen in the ionized gas is in O$^{2+}$ (as the O$^{2+}$ abundance is not explicitly computed).
Hydrogen recombination cooling and free-free cooling \citep{Tielens2005} are also included but only play minor
or negligible roles. We include Lyman $\alpha$ cooling and O\textsc{i} 6300\AA{} line cooling \citep{Tielens2005},
which can become important close to the H$^+$/H transition. In the neutral PDR gas, we included as main coolants the 
C$^+$ 158 $\mu$m, O 63 and 146 $\mu$m fine structure lines \citep{Rollig2006}, and the H$_2$ rotational \citep{Hollenbach1979} and 
vibrational \citep{Rollig2006} line cooling. For the deeper molecular regions, we include CO and $^{13}$CO (assuming a fixed 
$^{12}$CO/$^{13}$CO abundance ratio of 70, as isotopologue chemistry is not included) rotational cooling \citep{Hollenbach1979} and 
gas-grain heat transfer \citep{Hollenbach1989}. This last process requires computing an estimate of the dust temperature,
which is described in the next subsection.

\subsection{Radiative transfer}

The radiative transfer of EUV photons ($h\nu>13.6$ eV) is computed in a wavelength-dependent way.
The EUV field is mainly important in the ionized region. The dust size distribution and dust abundance
in H\textsc{ii} regions remain uncertain. While dust is observed in H\textsc{ii} regions, it is expected to 
be less abundant due to sublimation of the dust material by energetic photons and to its expulsion by
radiation pressure (see \citealt{Akimkin2015,Akimkin2017} for a study of the dynamics of dust grains 
in a pressure-driven H\textsc{ii} region). Here, we adopt a simplification similar to the model of 
\cite{Hosokawa2006b} and neglect the effect of dust in our calculation of EUV transfer, and extinction
only occurs through photoionization of H. The radiative transfer equation for the EUV
photon number flux $F_\mathrm{EUV}(r,\nu)$ is then
\begin{equation}
\frac{1}{r^\alpha}\frac{d \left( r^\alpha F_\mathrm{EUV}(r,\nu) \right)}{dr} = - n(\mathrm{H}) \sigma_\mathrm{H}(\nu) F_\mathrm{EUV}(r,\nu)
\end{equation}
where $\alpha$ is again the dimensionality-dependent exponent (0 for planar, 2 for spherically symmetric).
We use for the photoionization cross section $\sigma_\mathrm{H}(\nu)$ the values given in Table 7.2 of \cite{Tielens2005}.
During the chemistry stage, the photoionization rate and photoionization heating rate are then obtained by integrating 
over $F_\mathrm{EUV}(r,\nu)$.

In contrast, we do not compute the wavelength-dependent radiative transfer of FUV photons 
($h\nu<13.6\,\mathrm{eV}$).
We only consider the integrated radiation field over
the 912 - 2400 \AA{} domain, in units of the Habing standard interstellar radiation field:
\begin{equation}
G_0 = \frac{\int_{912\AA}^{2400\AA} u_\lambda(\lambda) d\lambda}{u_\mathrm{Habing}}
\end{equation}
where $u_\mathrm{Habing} = 5.337\times 10^{-14} \, \mathrm{erg\, cm}^{-3}$ \citep{Habing1968}. The radiative transfer
equation for $G_0$ is
\begin{equation}
\frac{1}{r^\alpha}\frac{d \left( r^\alpha G_0(r) \right)}{dr} = - n_\mathrm{H} \sigma_d G_0(r)
\end{equation}
where we adopt $\sigma_d = 1.2\times 10^{-21} \,\mathrm{cm}^{2}$ for the average FUV dust extinction cross section
per hydrogen nucleus.
In a first approach, we do not treat anisotropic scattering by dust grains, which would require a more complex radiative transfer calculation \citep{Flannery1980,Roberge1983,Goicoechea2007}.
We also assume dust to be fully coupled to the gas, so that $\sigma_d$ stays constant.
This $G_0$ value is then used for the computation of photoelectric heating, and of the dust temperature
following \cite{Hollenbach1991}.
The photodissociation of H$_2$ and CO occurs through line absorption and these species are thus capable of efficient
self/cross-shielding. We thus use the shielding approximations of \cite{Draine1996} for H$_2$ photodissociation 
and of \cite{Lee1996} for CO (which takes into account cross shielding of CO by H$_2$) to compute their 
photodissociation rates. All other photoreactions have rates of the form 
\begin{equation}
k = k_0\,G_0^\mathrm{edge} e^{-\beta\,A_\mathrm{V}}
\end{equation} 
where $k_0$ is the unshielded rate under $G_0=1$, and $G_0^\mathrm{edge}$ is the value of $G_0$ at the left edge 
of the grid (toward the star).

The external radiation field is assumed to come from a nearby star and is thus a beamed radiation
field. We use a blackbody radiation field that we parametrize by its effective temperature $T_*$ and
the FUV intensity reaching the edge of the grid $G_0^\mathrm{edge}$. We assume no extinction
between the star and the edge of the grid.

\subsection{Global time step determination}

The global time step is fixed by a combination of hydrodynamical
and photochemical conditions. The hydrodynamical stage requires
satisfying the usual CFL condition \citep{Courant1928}
\begin{equation}
\Delta t_\mathrm{hydro} = \min_i\left(  C \frac{\Delta r_i}{c_{s,i}}\right), \qquad C \le 1
\end{equation}
where $c_s$ is the sound speed, $i$ runs over grid cells and the Courant number $C$ has been set at 0.4 in the models
presented in this article. This constraint is complemented by an 
additional constraint designed to avoid the collapse
of our Lagrangian cells:
\begin{equation}
\Delta t_\mathrm{hydro} < \min_i \left(\frac{\Delta r_i}{\max(u_i,u_{i-1}) - \min(u_i, u_{i+1})}\right).
\end{equation}
During the thermochemical stage, the ODE solver automatically subdivides the time step as necessary. During this stage,
we thus only need to ensure that the variables that can affect the other stages (hydrodynamics and radiative
transfer) do not vary by too much during the full time step duration.
These variables are the thermal energy $e$ (which will affect the hydrodynamics),
and the number densities of the species involved in the radiative transfer : H, H$_2$ and CO.
For these variables, a timescale is estimated at the beginning of the time step as the ratio of the variable
to its variation rate, and the chemical time step $\Delta t_\mathrm{chem}$ is set at half the minimum of these timescales.
The global time step is finally taken as the minimum of the two values
\begin{equation}
\Delta t = \min \left( \Delta t_\mathrm{hydro}, \Delta t_\mathrm{chem} \right )
\end{equation}

\section{Code validation}\label{sec:Tests}

\subsection{Pure hydrodynamics test problems}

We validate the hydrodynamical solver of the \textit{Hydra} PDR code on two standard test problems:
the Sod shock tube \citep{Sod1978}, and the Sedov blast wave \citep{Sedov1959}. Both tests are performed with $\gamma = 1.4$.

\FigureSodShockTubeTest{}%

The Sod shock tube test problem consists of an initial configuration with different states on each side
of the midplane of the computational grid: $\rho_\mathrm{l} = 1$, $P_\mathrm{l} = 1$ and $u_\mathrm{l}=0$
in the left half region, and $\rho_\mathrm{r} = 0.125$, $P_\mathrm{r} = 0.1$ and $u_\mathrm{r}=0$ in the right
half. Because of the central discontinuity, this configuration induces the development of a shock wave, a contact
discontinuity and a rarefaction wave. We compute the evolution for a time $t=0.2$ on a 150-point grid covering
the range [0,1] in positions, and present the comparison of our results with the analytical solution on Fig. \ref{fig:SodShockTubeTest}. 
All three types of waves are correctly reproduced, with very sharp discontinuities ($\sim$2 grid points). 
The small discrepancies found at the contact discontinuity are due to initialization errors
(the initial discontinuity is not infinitely sharp due to the finite grid resolution) and are not smoothed out
during the evolution due to the total absence of numerical dissipation for pure advection in our 
Lagrangian scheme. The variations in grid spacing in regions that have been compressed or expanded
resulting from our Lagrangian scheme are clearly visible.

\FigureSedovPlanarBlastWave
\FigureSedovSphericalBlastWave

As a second test, we consider the Sedov blast wave problem: in a medium with zero initial pressure and velocity and $\rho_0 = 1$, energy
is deposited at a single point at the center of the domain, inducing the expansion of the central region and the
 propagation of a strong shock wave away from the center. For numerical reasons, 
the initial pressure is set at a negligible value of $10^{-30}$ rather than 0. We compute the evolution for a time $t=1$ on a 200-point
grid covering the range [0,1] in positions.
We consider this problem in both planar and spherical symmetry in order to test our solver in the different possible geometries. 
A comparison of our results with the analytical solution are shown on Fig. \ref{fig:SedovPlanarBlastWave} for the planar case and Fig. \ref{fig:SedovSphericalBlastWave} for the spherical case. Good agreement and sharp shock discontinuities ($\sim$ 2 grid cells) are again found in this strong
shock case.

\subsection{STARBENCH test : Pressure-driven expansion of an H\textsc{ii} region}

\FigureSTARBENCHearlyphase{}%
\FigureSTARBENCHlatephase{}%

As a test of the coupling between the hydrodynamics, chemistry and radiative transfer modules, we consider the test problem
of the pressure-driven expansion of a spherical H\textsc{ii} region. This problem has been proposed by \cite{Bisbas2015a} as
a standard benchmark for ionizing radiation hydrodynamics codes, comparing the results of numerical codes to various analytical prescriptions \citep{Spitzer1978,Hosokawa2006b,Raga2012a,Raga2012b}. The initial conditions of this test consist of a central star
emitting $10^{49}$ ionizing photons per second, in a homogeneous atomic medium with mass density 
$\rho_0 = 5.21\times 10^{-21}\,\mathrm{g\, cm}^{-3}$. The gas is assumed to be composed of hydrogen only.
An isothermal equation of state is used with $T=10^4\,\mathrm{K}$ in the ionized gas and either $T=100$ K or $T=1000$ K in the neutral gas.
We implement a pseudo-isothermal prescription using sharp ad hoc cooling functions. The recombination coefficient and photoionization cross
section used in this test are $\alpha_B = 2.7\times10^{-13}\,\mathrm{cm}^3\,\mathrm{s}^{-1}$ and $\bar{\sigma} = 6.3\times 10^{-18}\,\mathrm{cm}^2$.

In order to test both the early expansion behavior and the saturation of the H\textsc{ii} region radius at late time, two models were considered in \cite{Bisbas2015a}: an early phase model with neutral gas temperature $T=100$ K and run until $t=0.141$ Myr on a grid reaching $R_\mathrm{max}\simeq1.3$ pc, and a late phase model with neutral gas temperature $T=1000$ K and run until $t=3$ Myr on a grid reaching $R_\mathrm{max}\simeq6$ pc. In both cases, we use a grid of 1000 points.
We show the comparison of our results with the reference analytical prescriptions considered in \cite{Bisbas2015a} in Fig. \ref{fig:STARBENCHearlyphase} for the early phase model and in Fig. \ref{fig:STARBENCHlatephase} for the late phase model. In the early phase case, we find our results to follow the Spitzer formula \citep{Spitzer1978} at early times before switching to the Hosokawa-Inutsuka formula \citep{Hosokawa2006b}, similar to what is found with the multiple numerical codes tested in \cite{Bisbas2015a}. In the late case scenario, our results are intermediate between the Raga I \citep{Raga2012a} and Raga II \citep{Raga2012b} solutions, and follow very closely the time dependent linear combination of these two solutions proposed in \cite{Bisbas2015a} based on the numerical results of the tested codes (noted as STARBENCH on Fig. \ref{fig:STARBENCHlatephase}). In both cases, we thus find similar results to other established ionizing radiation hydrodynamics codes.


\section{Results}\label{sec:Results}

\FigureEvapDrawing

As a first application of this code, we investigate the structure of a photoevaporating PDR front.
We consider the edge of a molecular cloud exposed to stellar irradiation and photoevaporating freely into a low pressure medium.
This represents the phase where the H\textsc{ii} region surrounding the star has broken out of its parental molecular cloud and into the surrounding diffuse medium, and the edges of the remnants of the molecular cloud are being photoevaporated. As we focus on the local structure of the photoevaporation front, we consider a plane parallel geometry.
A schematic view of this photoevaporation scenario is presented on Fig. \ref{fig:EvapDrawing}.

The initial conditions consist of a uniform molecular medium (all hydrogen in H$_2$, 90\% of carbon in CO with the rest in atomic carbon) at an initial temperature of 20 K. In order to allow photoevaporation of the gas towards the star, the boundary on the left side (towards the star) of the computational domain is a free-moving fixed-pressure boundary. We chose for the external pressure on this boundary $P_\mathrm{ext}=10^4\,\mathrm{K\, cm}^{-3}$, representative of the diffuse medium surrounding the star forming cloud, and into which the gas from the H\textsc{ii} region is free to evaporate. The right boundary (towards the molecular cloud) uses a fixed wall boundary condition, but the final state of the run is taken so that no pressure wave has reached this outer boundary, so that its choice does not affect the results.

In order to study the physical structure of a photoevaporating PDR front and its variation with the environmental parameters, we ran a grid of models with varying initial gas densities and impinging radiation fields. 
The initial gas densities range from $10^2$ cm$^{-3}$ to $10^6$ cm$^{-3}$ and the impinging radiation field is parametrized by its FUV intensity $G_0$, with values ranging from $10^1$ to $10^5$.
To compare the impact of stellar type on the photoevaporation process, we consider two star types in our model grid: either a O star, for which we take as an example the O7Vp star $\theta^1$ Ori C, the main star of the trapezium illuminating the Orion Bar ($T_\mathrm{eff}=4\times10^4\,\mathrm{K}$, $L = 2\times 10^5 L_\odot$, \citealt{Simon-Diaz2006}), or a B star, for which we take as an example the Ae/Be star HD200775 illuminating the PDRs of NGC~7023 ($T_\mathrm{eff} = 1.9\times10^4\,\mathrm{K}$, $L = 1.5\times 10^4 L_\odot$ including both components A and B, \citealt{Alecian2008}).
We thus ran two grids of 25 models (5 density values and 5 $G_0$ values) corresponding to the two star cases.

In our plane-parallel configuration, a Str\"omgren-like distance can be defined as $d_\mathrm{Str\ddot{o}mgren} = \frac{J}{\alpha_B n^2}$ (where $J$ is the ionizing photon number flux), corresponding to the equilibrium thickness of the H\textsc{ii} region if the dynamics of the gas were frozen. 
In order to both allow sufficient space for the development and propagation of the photoevaporation front and still resolve sufficiently its initial formation with a 5000-points grid, we chose the size of our computational grid to be $200\times d_\mathrm{Str\ddot{o}mgren}$. In the cases where this is not sufficient to represent a meaningful column density of gas, we modify the grid size to correspond to a minimum total $A_\mathrm{V}$ value, which is chosen depending on the model gas density ($A_\mathrm{V}^{(\mathrm{min})} = 1, 1.5, 2, 3, 4$ for $n_0 = 10^2, 10^3, 10^4, 10^5$ and $10^6\,\mathrm{cm}^{-3}$ respectively). The final time for each run was chosen to allow propagation of the photoevaporation front up to roughly 90\% of the grid size, using an estimate of the front velocity based on an initial coarser grid of models. Finally, as the plane parallel geometry is only a good approximation if the sizes considered are not larger than the distance from the front to the illuminating star, we excluded the $n_0-G_0$ pairs that led to $d_\mathrm{Str\ddot{o}mgren}>d_\mathrm{star}/4$ (these pairs correspond to low density high illumination models).
 
One complication arises from the fact that in this 1D plane-parallel geometry, all the photoevaporated material remains on the line of sight to the star, progressively extinguishing its radiation field. Absorption of ionizing photons in the ionized region is proportional to the recombination rate and thus becomes insignificant at low densities. Only a limited region close to the ionization front can maintain a moderate density and contribute significantly to EUV extinction (cf. the ionized boundary layer in \citealt{Bertoldi1989}). This is however more problematic for its impact on the FUV transfer through dust extinction, the visual extinction from the star to the evaporation front increasing proportionally to the quantity of photoevaporated gas (as the dust to gas ratio is kept constant in our model). This prevents the existence of a stationary state as the FUV flux reaching the front decreases with time, and can also artificially modify the ratio of EUV to FUV photon fluxes reaching the photoevaporation front. An exact solution of this problem would require a full 3D simulation to allow the gas photoevaporated from the dense PDRs to escape out of the line of sight and into the diffuse surrounding medium. A full treatment of dust dynamics and destruction in the H\textsc{ii} region could also reduce the problem by strongly decreasing the extinction in the ionized gas.
As a first step, we chose here to further limit the run duration so that the extinction from the star to the ionization front remains lower than $A_\mathrm{V}=1$. This limitation only affected 3 of our models.

In each of our models, we define the ionization front as the position where 50\% of hydrogen is in H$^+$ and the dissociation front as the position where 50\% of hydrogen is in molecular form. The shock fronts are detected using the same shock detection algorithm that is used in the hydrodynamics solver to trigger flattening of the interpolation function \citep{Colella1984}.
In the following, we abbreviate ionization front as IF, dissociation front as DF and shock front as SF. 
 
\subsection{Photoevaporation : examples and types}

\subsubsection{A typical example model}\label{sec:ExampleModelOne}

\FigureExampleRunOneFronts{} %

\FigureExampleRunOne{} %

We first present the results of an example model with $n_0 = 10^5$ cm$^{-3}$ and $G_0=10^4$ and $T_* = 4\times 10^4$ K.
The evolution of the IF, DF and SF positions is shown on Fig. \ref{fig:ExampleRunOneFronts}.
At first occurs a very short initial phase during which the IF and the DF quickly propagate (as the gas is being ionized/dissociated with negligible recombination/reformation), while the gas does not have time to move (R-type phase, cf. \citealt{Spitzer1978} and \citealt{Draine2011} for a detailed discussion of this R-type/D-type distinction). As they slow down when approaching what would be their equilibrium position if the gas was kept static, the gas heated by photoionization heating and H$_2$ UV pumping heating starts to expand, and the fronts become D-type and are preceded by a shock front while propagating much more slowly relative to the gas. The initial R-type phase is too short to be visible on the figures presented here. 
While the IF and the DF are initially each preceded by their own shock front, we see that the IF quickly catches up with the DF (and their shock fronts merge), after which they remain at a roughly constant distance, preceded by a single shock front. After the initial R-type phase, the R-type/D-type transitions and the transitory perturbations that follow the merging of the two shock fronts, a quasi-stationary regime seems to be reached after a few thousand years, in which the front properties vary due to the progressive extinction of the UV field by the increasing column of photoevaporated gas.

\FigureScenarioMap{} %

Figure \ref{fig:ExampleRunOne} shows the profile of the main physical and chemical variables at different times during the propagation of the fronts. The left panel shows the profiles at five different times, while the right panel shows a zoom on the profile at the final time. The shock front induces the formation of a compressed layer where the density reaches up to $7\times 10^6$ cm$^{-3}$, a factor of $70$ above the initial cloud density (cf. top panel). This shock is driven by the large pressure difference between the unperturbed cloud, and the compressed layer and ionized region (third panel). We see that the pressure in the H\textsc{ii} region does not equilibrate with the low external pressure imposed at the moving boundary ($10^4\,\mathrm{K\, cm}^{-3}$). This is due to the slightly supersonic velocity of the photoevaporation flow, which does not allow a rarefaction wave to move towards the photoevaporation front. As a result, this external pressure has a negligible effect on the results as long as it is very small compared to the IF and PDR pressures. We however note a clear pressure difference between the atomic and molecular compressed layer on one hand, and the ionized region on the other hand, which is the consequence of the photoevaporation process: the expansion of the photoevaporating gas towards the star exerts by reaction a force on the neutral layer. This differs from expanding (confined) H\textsc{ii} regions simulations (e.g. \citealt{Hosokawa2006b}), where the compressed PDR is found to be at pressure equilibrium with the ionized region. In the final state shown here ($t=5048\,\mathrm{yr}$), the pressure is $P\sim 10^8 \,\mathrm{K\, cm}^{-3}$ in the atomic and molecular PDR, while it is $\sim 4\times 10^7 \,\mathrm{K\, cm}^{-3}$ in the ionized region. In addition, we see that after the initial transitory regime that follows the merging of the IF-induced and DF-induced shocks, the pressure is roughly constant across the molecular part of the compressed layer (except close to the shock) and the higher density part of the atomic region. Pressure gradients are only found close to the IF, and in the coldest, most inner part of the compressed layer. This quasi-isobaric profile results in a strong density gradient with density increasing from the IF to the shock front, following the temperature gradient induced by photoelectric heating in the neutral region. We also notice a temperature spike in the atomic region, just before the DF, which we found to be caused by H$_2$ UV pumping heating of the gas in the region where H$_2$ has not yet self-shielded. The ionized photoevaporation flow is found to have a velocity of $\sim 10\,\mathrm{km/s}$ relative to the molecular cloud (close to the sound speed in the ionized gas), consistent with what has been observed in the face-on surface of the Orion Nebula \citep{Goicoechea2015}. Finally, the shock front propagates at $\sim 3$ km/s, inducing a very thin $300-400$ K layer at the back of the compressed layer. Behind the compressed layer, the gas, while unaffected dynamically, is heated by photoelectric heating from the fraction of FUV photons that manage to cross the compressed layer.

In the following, we will focus on the regions that are relevant to PDR observations: we define the neutral atomic layer (AL) as the layer between the IF and the DF, and the warm molecular layer (WML) of the compressed PDR as the layer between the DF and the position where the temperature has fallen to 100 K (or the shock front if all the compressed layer remains above 100 K). This warm molecular layer is the region of emission of most of the molecular PDR tracers (rotational H$_2$ emission, reactive ions and radicals such as CH$^+$, SH$^+$, HOC$^+$ and OH, high-$J$ CO rotational emission, etc., see e.g., \citealt{Allers2005,Agundez2010,Goicoechea2011,Nagy2013,Joblininprep}). The positions of these layers are indicated in Fig. \ref{fig:ExampleRunOne} (right panel, in the pressure plot).

\subsubsection{Photoevaporation types}

\FigureExampleRunTwoFronts{} %

\FigureExampleRunTwo{} %

We find that, in our grid of models, not all models behave qualitatively like the example model of the previous subsection. We can distinguish three main types based on the behavior of the IF and DF. 

i) In the first one, the photoevaporation at the IF and DF cannot produce a sufficient pressure to confine the cloud. This happens mainly for high initial density models, and the cloud freely expands into the lower pressure external medium, with the expansion flow sweeping away both the IF and the DF. We call this a "no front" evaporation scenario, as there is no photoevaporation front exerting pressure on the cloud, and evaporation simply occurs as a wide rarefaction wave propagating into the cloud, with the radiation field playing a negligible role. In a more realistic setting, the swept away IF and DF would at some point (having moved towards the star) reach a position were they are submitted to a radiation field high enough that photoevaporation can generate sufficient pressure again. A photoevaporation front would thus finally form closer to the star and submitted a stronger radiation field. It would thus correspond to a higher $G_0$ model in our grid.

ii) In the second case, the IF does not generate sufficient pressure and is swept away, but photoevaporation of neutral gas at the DF is sufficient to keep it contained and to drive a shock wave into the cloud. We call this case a "DF-contained" evaporation scenario, as it is only photoevaporation at the DF that can contain and compress the cloud. As an example model exhibiting this behavior, we show here the model with $n_0 = 10^5$ cm$^{-3}$ and $G_0=10^4$ and $T_* = 1.9\times 10^4$ K. Figure \ref{fig:ExampleRunTwoFronts} presents the evolution of the DF and SF positions. The IF does not appear on the figure, as it is immediately swept away towards negative position values. The DF and an associated SF however propagate into the cloud, with a wide compressed layer and a neutral atomic photoevaporation flow. Evidence for such a neutral photoevaporating flow might be observable in the kinematics of C$^+$ emission. The profiles of the main variables are presented at different times on Fig. \ref{fig:ExampleRunTwo}. We see in this case a much smoother pressure gradient extending from the atomic region to the shock front. Indeed, photoelectric heating plays here a major role in the photoevaporation and plays over a wider region than the photoionization heating that dominated in the previous example of Sect. \ref{sec:ExampleModelOne}. This also translates into a continuous velocity gradient, in contrast to the sharp velocity jump around the IF in the previous example. The smaller pressure difference between the compressed layer ($\sim 4\times 10^7\,\mathrm{K\, cm}^{-3}$) and the unperturbed cloud also results in a lower shock velocity ($\sim 1.5$ km/s) heating the gas to less than 200 K only.

iii) The third case corresponds to the scenario occurring in the first example model (Sec. \ref{sec:ExampleModelOne}), where photoevaporation at the IF is sufficient to compress the cloud, and a IF-DF-SF triplet jointly propagate into the cloud. We thus call this case an "IF-contained" evaporation scenario.

The behavior of each model in terms of these three evaporation scenarios is presented on Fig.\ref{fig:ScenarioMap}, where the three scenarios are color-coded as yellow for IF-contained, green for DF-contained, and blue for no-front. The left panel shows the models for an O-type star, and the right panel for a B-type star. The white upper left region in each panel corresponds to the models that have been excluded because the Str\"omgren scale was too large compared to the distance to the star for our plane-parallel geometry to be relevant. For these models, the front quickly moves to large distances from the star (during the R-type phase), where the radiation field is decreased. After the transition to the D-type phase, these situations would thus correspond to lower $G_0$ models in our grid.
With both stellar types, high-density low-$G_0$ models show no-front evaporation, as the dense cloud naturally tends to expand into the low pressure surrounding medium while the low radiation field allows only very weak photoevaporation, insufficient to contain the expansion of the cloud. This no-front region is smaller in the O-star models, because at constant $G_0$, the higher ionizing photon flux of the O-star results in more efficient photoevaporation. The DF-contained cases occupy a intermediate region between the no-front and IF-contained domains, with a larger extension for higher initial density, and for B-star models compared to O-star models.

\subsection{Photoevaporation : PDR compression}

We now investigate the consequences of photoevaporative compression on the structure of PDRs and in particular on the properties of the warm molecular layer of the PDR. 

\subsubsection{P-G$_0$ relation}

\FigurePGrelation{}%

A correlation between the pressure derived from warm molecular tracers in a PDR and the FUV intensity $G_0$ that illuminates it has been derived from PDR observations \citep{Chevance2016,Joblininprep,Wuinprep}. It has been proposed to be evidence for photoevaporation being the main mechanism inducing the high gas pressures derived from excited molecular tracers.
We thus investigate here how the pressure in the warm molecular layer correlates with the $G_0$ value of the FUV field reaching the dissociation front in our photoevaporating PDR models.

\FigurePressureVariations{} %

We take here the average thermal pressure in the warm molecular layer, and the $G_0$ measured just before the dissociation front (we take the point where 20\% of hydrogen nuclei are in the form of molecular hydrogen). We show these values for the final state of our models in Fig. \ref{fig:PGrelation}, distinguishing O-star (red) and B-star (blue) models, and models exhibiting IF-contained behavior (filled circles) and DF-contained behavior (open circles). Models in the no-front domain are not represented as no significant compression of the PDR occurs. Two of the B-star models ($n_0 = 10^2$ cm$^{-2}$, $G_0 = 10^2$ and $n_0 = 10^4$ cm$^{-2}$, $G_0 = 10^5$) do not appear on this figure as their compressed layer is still fully atomic at the final time of the model, so that they have no warm molecular layer according to our definition.
 
Figure \ref{fig:PGrelation} shows that we find a clear correlation between $P$ and $G_0$ in our models. At first order, the relation seems to be linear, with some deviation at the highest $G_0$ values. Most of the points are found to be within a factor of 4 from $P/G_0 = 2\times 10^4$ (the dotted lines on Fig. \ref{fig:PGrelation} show $P/G_0 = 5\times 10^3$ and $P/G_0 = 8\times 10^4$). This is compatible with the values derived from the observations in \cite{Joblininprep}. \cite{Chevance2016} also found $P/G_0 = 10^3-10^4$ from observations of several star forming regions in the Large Magellanic Cloud and Small Magellanic Cloud. Our photoevaporating PDR models thus naturally predict a $P-G_0$ correlation quantitatively similar to the observed correlation, almost independently of the initial density of the cloud (if the conditions allow a photoevaporation front to exist).

In addition, we find a systematic shift by a factor of $\sim4$ for the $P/G_0$ ratio between the O-star models and the B-star models. Models with higher photoionizing photon fluxes thus seem to show more efficient compression at equal FUV illumination.
\cite{Joblininprep} found no apparent difference between B-star PDRs (in NGC~7023) and O-star PDRs, but did not quantify the uncertainties on the derived $P$ and $G_0$ values.
Given the low number of observational points in \cite{Joblininprep} and the probably large uncertainties, it is unclear whether this systematic shift seen on Fig. \ref{fig:PGrelation} is contradicted by the observations.
In addition, we find no clear difference between DF-contained and IF-contained cases in term of the $P-G_0$ ratio at the final time of the runs.
A more detailed observational investigation of the $P-G_0$ correlation including a larger number of PDRs around various stellar types would thus be necessary to better constrain the relative importance of ionizing and non-ionizing UV photons in the photoevaporative compression process.


\subsubsection{Pressure gradients}

Isobaric (stationary) PDR models have been found to give better results than constant density models for explaining excited molecular tracers \citep{Allers2005,Nagy2013,Cuadrado2015,Joblininprep}. We thus investigate here to which extent an isobaric description of the warm molecular layer where these tracers are emitted is adequate. We estimate the total pressure variation occurring in the WML relative to its average pressure, $\Delta P/P$. As the pressure variation simply defined as $P(b)-P(a)$ (where $a$ and $b$ are the boundaries of the WML) can be very sensitive to the way we choose the boundaries if there are sharp variations at the boundary (e.g. the shock front), we choose another estimate, more representative of the bulk of the WML: $\Delta P = \mathrm{median}(\frac{dP}{dx})_\mathrm{WML} \times d_\mathrm{WML}$, where $d_\mathrm{WML}$ is the thickness of the warm molecular layer, and $\mathrm{median}(\frac{dP}{dx})_\mathrm{WML}$ is the median pressure gradient in the WML. Finally, as some of the models show small numerical oscillations close to the DF, we average the result over the last 5\% of the run duration to smooth out their impact.

The results are presented on Fig. \ref{fig:PressureVariations}, with O-star models on the left panel and B-star models on the right panel. The pressure is always found to be increasing from the DF towards the shock (positive values of $\Delta P$). Excluded models and models with no-front evaporation are left in white. Models where the compressed shell remains fully atomic are also left in white, as the warm molecular region is not defined (two B-star models with $n_0 = 10^2$ cm$^{-2}$, $G_0 = 10^2$ and $n_0 = 10^4$ cm$^{-2}$, $G_0 = 10^5$).  More generally, low-density high-$G_0$ models have a DF that propagates very far from the IF during the initial R phase. During the D phase, the IF-induced shock thus has to compress a large quantity of gas before catching up with the DF. As a result, the warm molecular layer, when it exists, tends to be much smaller than in other models, resulting in very low $\Delta P/P$ as seen on models on the uppermost diagonal of Fig. \ref{fig:PressureVariations}.
Pressure variations in the WML seem to be larger for B-star models than for O-star models. Overall, pressure variations in the WML are found to be at most of a factor of $\sim2$, with most O-star models having variations by less than $\sim60\%$. Low $G_0$ models have very low pressure variations, as the low UV field become insufficient to warm a wide layer of molecular gas.

\FigureDensityRatio{} %

While moderate pressure gradients are present, a constant pressure approximation seems to be a more reasonable first order approximation than a constant density representation (which results in pressure relative variations equal to the temperature relative variations across the PDR). In addition, we note that the pressure gradients are found to be increasing towards the inside of the PDR, which is the opposite of what results from constant density PDR models. We thus conclude that isobaric PDR models are a better approximation than constant density models to represent photoevaporating PDRs.

\subsubsection{Density contrasts}

The [C\textsc{ii}] 158 $\mu$m and [O\textsc{i}] 63 $\mu$m lines, which are often used to constrain the PDR conditions, have a significant fraction of their emission originating in the atomic layer of the PDR. Due to the large density contrasts between the atomic layer and the warm molecular layer, this can lead to problems when trying to interpret both these lines and warm molecular tracers with constant density models. We thus present here the density contrasts between the AL and the WML found in our photoevaporating models. We take as an indicator the ratio of the average gas densities in each layer.

The results are presented on Fig. \ref{fig:DensityRatio}. For O-star models, we find a density contrast between one and two orders of magnitude for all model, with the smallest values for high-density high-$G_0$ models. The density contrasts are a bit lower for B-star models, with the lowest contrasts being a factor $\sim 4$ found for moderate densities ($10^3-10^4$ cm$^{-3}$) and high radiation fields ($10^3-10^4$). Overall, we thus find density ratios between the atomic layer and the warm molecular layer from a few $10^{-2}$ to a few $10^{-1}$. This is similar to the density contrasts usually considered in the clumpy PDR hypothesis. The natural density gradient of a photoevaporating PDR could thus by itself explain the density contrasts that are necessary to explain the excitation of the observed tracers, without needing a small scale clumpiness hypothesis. This does not exclude the presence of substructures in PDRs (which might originate from pre-existing density fluctuations in the molecular cloud), but shows that their existence is not implied by the density contrasts derived from different tracers.

\subsubsection{Sizes of the different layers}

\FigureWMLsize{} %
\FigureSizeRatio{} %

We now investigate the thickness of these different layers. We first present the thickness of the WML on Fig. \ref{fig:WMLsize}. To first order, the thickness is found to increase with decreasing density and $G_0$, as expected from the fact that the lower pressure generated by photoevaporation under lower radiation field induces a weaker density increase, and thus results in more extended structures. The WML thickness is found to be lower for O-star models (from a few 0.1 mpc to a few mpc), than for B-star models (up to a few 10 mpc), as expected from the higher pressures found in O-star models. The sizes found for the high radiation fields ($G_0 = 10^3 - 10^4$) and initial gas densities ($n_0 = 10^4-10^5\,\mathrm{cm}^{-3}$) that are typical of high excitation galactic PDRs correspond to a few mpc. This is consistent with the few arcsec layer that has been observed from the most excited molecular tracers in PDRs such as the Orion Bar \citep{Allers2005,Goicoechea2016,Goicoechea2017} or NGC~7023~NW \citep{Fuente1996,Lemaire1996}, as 1" corresponds to $1.9$ mpc at a 400 pc distance.
The predicted sizes are however at or beyond the limit of the angular resolution of most telescopes, even for relatively close-by PDRs. Further ALMA observations and future JWST observations in a larger number of PDRs will thus be crucial to resolve the spatial structure of PDRs and constrain our dynamical PDR models.

We also find the thickness of the atomic layer to be systematically larger than the warm molecular layer. The thickness ratios of the AL and WML are presented on Fig. \ref{fig:SizeRatio}. Thickness ratios range from 1 to 30, with values for the $G_0 = 10^3 - 10^4$ range typically ranging from 2 to 8. \cite{Goicoechea2016} observed that the atomic layer in the Orion Bar was significantly more extended than the warm molecular layer, with a thickness ratio of $\sim 7$ (15" for the atomic layer and $\sim 2"$ for the compressed warm molecular layer). Although we do not quantitatively reproduce these absolute sizes for the physical parameters of the Orion Bar, our models show that photoevaporating PDRs can at least qualitatively explain the observed extended atomic layer. Further exploring the impact of crucial PDR processes such as the photoelectric efficiency, H$_2$ formation efficiency, and dust extinction properties and refining their treatment could help obtaining a better quantitative match.

\subsubsection{Shock velocities}

\FigureShockVelocities{} %

Finally, all of our photoevaporating models result in a low velocity shock preceding the IF and DF. The shock velocities are presented on Fig. \ref{fig:ShockVelocities}. The shock velocities in our models range from 0.5 to 7 km/s and seem to be controlled mainly by the $G_0/n_0$ ratio. The shock velocity are also systematically lower for B-star model than for O-star models, consistently with the lower pressures and lower compressions discussed above.

While these low velocity shocks result in the presence of a shock-heated layer at the back of the compressed layer (with temperatures of a few 100 K), this layer is found to be much thinner than the warm molecular layer of the PDR. As the warm molecular layer of the PDR has temperatures that are similar and higher, and contains a larger column density of gas,
it is unlikely that thermal excitation from the shock heating could in general be distinguished observationally from the PDR emission, unless having sufficient angular resolution to resolve the spatial separation between the two layers. 
Specific tracers of the shock chemistry might however be investigated, although the shock velocities are insufficient to induce significant dissociation. The chemical conditions in the shock heated gas might differ from the warm molecular layer of the PDR as the UV illumination reaching the shock heated layer is much lower. Such chemical differentiation will be investigated in future works. In addition, grain shattering could occur at these low shock velocities \citep{Tielens1994}, causing a difference in dust extinction properties between the PDR and the molecular cloud.
The spatial separation between the shock and the PDR is found to increase with time (cf. Fig. \ref{fig:ExampleRunTwoFronts}), with a faster increase for B-star models. A more careful investigation of the dependency of this increase to the initial conditions would however be necessary to discuss quantitatively the use of this separation as an age tracer.

\section{Discussion}\label{sec:Discussion}

We discuss here the main limitations of our model, and the way they could be addressed in future works.

\subsection{Impact of magnetic fields}

Our model neglects the effect of magnetic fields on the dynamics of the gas. It is thus equivalent to assuming the magnetic field to be parallel to the radiation field illuminating the cloud. In the case where the magnetic field is perpendicular to the radiation field, it would be enhanced by the compression of the gas, and thus tend to resist this compression through an enhanced magnetic pressure. While an exact description of this effect would require including the magneto-hydrodynamics of our partially ionized gas, we discuss here the likely effects through which magnetic fields could modify our results.

\cite{Motoyama2013} studied the impact of a preexisting magnetic field on the photoevaporative compression of bright-rimmed clouds. They found that the magnetic field can have a non negligible impact on the density profile of the compressed cloud. Their model however described very cold PDRs (gas temperature of 30 K) and might thus overestimate the compression in the PDR layer and the importance of magnetic pressure relative to thermal pressure.
If we consider the extreme case where the magnetic field scales as $B\propto \rho$ (magnetic field perpendicular to the radiation field and frozen in the fluid) during the planar compression induced by the photoevaporation, we see that while the magnetic field might be significantly enhanced in the coldest part of the compressed layer immediately after the shock (where compression factors of the order $10^1 - 10^2$ are observed in most models), the density decreases back to lower values in the warm atomic and molecular layers of the PDR, with densities in the atomic layer that can be lower than the initial density of the cloud. Given the strongly enhanced thermal pressure in these regions, it is likely that it dominates there over magnetic pressure, assuming that the magnetic field was initially at equipartition with the thermal pressure in the unperturbed molecular cloud. Magnetic pressure could however play an important role in the deeper region of the compressed layer, reducing the density increase.

\cite{Henney2009} have carried out 3D MHD simulations of the photoionization of a magnetized molecular globule. They found that the initial magnetic pressure needs to be $\sim100$ times greater than the thermal pressure in the initial globule in order for the magnetic field to have a strong effect on the dynamics of the compression of the globule. For lower magnetic fields, the enhancement of the thermal pressure by radiative heating makes it dominant over magnetic pressure. \cite{Mackey2011} reached a similar conclusion in their 3D MHD simulation of the photoionization of pillars and globules at the edge of an H\textsc{ii} region.

\subsection{Dust evolution in the ionized region and PDR}

In our model, we have considered simple dust properties with constant extinction properties in the gas, but neglected the effect of dust extinction on ionizing photons. A more complete dust treatment should include several important effect that can cause an evolution of the dust optical properties.
\cite{Akimkin2015, Akimkin2017} have studied the effect of the dynamical decoupling of dust and gas in a 1D hydrodynamical model of an expanding H\textsc{ii} region. They find that the drift of grains relative to the gas due to radiative pressure results in the complete removal of big grains and a significant removal of PAHs from the H\textsc{ii} region. They find a global decrease of the dust-to-gas ratio in  the H\textsc{ii} region by a factor between 2 and 10.
This segregation effect would significantly reduce the dust extinction in the ionized gas, and modify the dust extinction curve by changing the dust size distribution inside the H\textsc{ii} region.
In addition, the radiation field of the star will destroy \citep{Jones2004,Guhathakurta1989} and process \citep{Compiegne2008} the dust grains, further reducing the extinction.

In addition to controlling how much dust remains in the H\textsc{ii} region, any evolution of the dust optical properties can strongly affect the structure of the PDR itself. \cite{Goicoechea2007} have shown that variations of the dust size distribution across the PDR could result in order-of-magnitude differences in the amount of UV photons reaching the deeper layers of the PDR.
While dust populations inside dense molecular clouds are expected to be in large part composed of large grains due to coagulation, dust emission in PDRs shows the presence of a population of (very) small grains and PAHs. The processing of grains into progressively smaller sized grains (and then PAHs) has been proposed to be linked to the impact of the UV field on dust \citep{Rapacioli2005,Berne2007,Pilleri2012}.  
Grain shattering in the shock front could also play a role in transforming a predominantly large grain population in the initial molecular cloud into wider size distribution enriched in (very) small grains.
In addition to affecting the extinction of the UV radiation field, the photoprocessing of dust grains might also release small hydrocarbons \citep{Alata2014,Alata2015} or H$_2$ molecules \citep{Jones2015} into the gas.
A full treatment will require to follow the evolution of the dust properties in a position-dependent and time-dependent manner. 
We will explore these multiple effects of the time-dependent evolution, destruction and drift of dust grains in a future study.

\subsection{Evaporated gas on the line of sight}

One of the main drawbacks of our 1D plane parallel geometry is that all the photoevaporated gas accumulates on the line of sight between the star and the cloud, resulting in a growing column density of very low density ionized gas in which dust extinction can significantly reduce the UV radiation field reaching the IF and DF. Dust expulsion from and photodestruction in the H\textsc{ii} region could in part limit the effect of the accumulated gas by reducing its extinction. In a realistic configuration, the main mechanism that will limit the amount of ionized gas on the line of sight will be its escape from the cavity surrounding the star and into the diffuse medium surrounding the molecular cloud. \cite{Henney2005} have simulated the flow of ionized gas escaping from a blister H\textsc{ii} region in a 2D geometry. In their models (which correspond roughly to $n_0=10^6\,\mathrm{cm}^{-3}$, $G_0$ of a few $10^4$ for our O-star models), they find that the flow reaches a quasi-steady state with a profile that depends mainly on the curvature of the IF and the distance to the star. The amount of gas between the star and the front does not increase during this phase as the gas can be swept away by diverging sideways and by moving past the star. A first step to improve our 1D model would thus be to limit the effective column of ionized gas extinguishing the radiation field based on such 2D simulation results.

\subsection{PDR physics and chemistry}

As a first step towards the modeling of photoevaporating PDRs, the model presented in this paper uses simplified approximations for several of the physical and chemical processes. Although these approximations are sufficient to obtain a reasonable description of the heating and cooling in the PDR, and thus of its temperature profile, a quantitative comparison of our model to specific tracers will require a more detailed treatment of the chemistry and excitation of these tracers. Moreover, the efficiency of crucial processes such as the photoelectric heating efficiency or the H$_2$ formation efficiency could quantitatively modify the results presented here, although we expect our conclusions to be qualitatively unaffected. The efficiency of both these processes however remain uncertain. 

The H$_2$ formation efficiency (see \citealt{Wakelam2017} for a comprehensive review) was for instance shown to be higher in PDRs than in the diffuse medium \citep{Habart2004}, with variations depending on the local conditions. Several H$_2$ formation models have been proposed to explain this increased efficiency, involving chemisorbed H atoms \citep{LeBourlot2012,Cazaux2004} and/or UV-induced temperature fluctuations of small grains \citep{Bron2014}.
As an example, we compared the results using a H$_2$ formation efficiency of $3\times10^{-17}$ and $1\times 10^{-16} \,\mathrm{cm}^3\,\mathrm{s}^{-1}$ for the example model of Sect. \ref{sec:ExampleModelOne}. The higher formation efficiency is found to increase the temperature in the atomic region by up to 50\% and reduce its density accordingly. This only results in a marginal increase in pressure in the compressed layer and in shock velocity.
The impact of each of the major processes will be further investigated in future developments of the code.

As the focus of this paper was on the physical structure of the PDR resulting from photoevaporation, we did not discuss the impact of this dynamics on the chemistry.
These aspects will be investigated in a future paper. The propagation of the dissociation front through the molecular gas can induce important non-stationary chemical and excitation 
effects \citep{Storzer1998}, which might be crucial to explain specific observational tracers of the PDR. In addition, ice-covered dust grains from the molecular cloud passing through the shock front and the progressive UV illumination will release the chemical products of dark cloud ice chemistry into the gas phase (cf. e.g. \citealt{Kirsanova2009} for numerical simulations of this effect around a pressure-driven expanding H\textsc{ii} region).

\section{Summary and conclusions}\label{sec:Conclusion}

We have developed a 1D hydrodynamical PDR code, coupling 1D hydrodynamics, EUV and FUV radiative transfer and time-dependent thermo-chemical evolution.
We validated this code on standard test problems in pure hydrodynamics and ionizing-radiation hydrodynamics.
As a first application, we simulated the photoevaporation of the edge of a molecular cloud exposed to the radiation field of a nearby massive star, where the gas 
is free to evaporate into a low-pressure medium. We investigated the morphology of the resulting photoevaporation front across a large range of physical parameters, 
in particular its pressure and density structure in relation to recent observational results in PDRs. Our main results are:
\begin{enumerate}
\item Photoevaporation at the ionization front and dissociation front generates high pressures in the PDR, compatible with those deduced from recent observations.
\item This results in a $P-G_0$ correlation with a roughly linear behavior, similar to the correlation that has been deduced from observations. In addition, we find a systematic shift of the relation with stellar type, with higher $P/G_0$ ratios for hotter stars.
\item The pressure inside the molecular layer is found to vary by at most a factor 2, with a gradient increasing in the direction opposite to the star. This could explain the better success recently reported for constant-pressure PDR models compared to constant-density models to explain warm molecular tracers (e.g. H$_2$, CH$^+$, high-$J$ CO lines,...) in PDRs.
\item The density is found to be 1 to 2 orders of magnitude lower in the atomic layer than in the warm molecular layer of the PDR. This strong density gradient is consistent with the contrasting densities that have been derived from the observations of different tracers. This shows that although pre-existing density inhomogeneities might be present in PDRs, the observed density contrasts between tracers do not imply small-scale clumpiness.
\item The photoevaporation front is preceded by a low velocity shock (0.5 to 7 km/s) which compresses the molecular gas before it enters the PDR (compression factors ranging from a few to several 100 depending on the conditions).
\end{enumerate}

Our results highlight the need for further high angular and high spectral resolution observational investigation of several aspects of PDRs in order to better constrain our dynamical PDR models.
The $P-G_0$ correlation can give strong constraints on the details of the photoevaporation process (e.g. relative importance of ionizing vs. non-ionizing UV photons),
but is observationally based on a very limited number of PDRs. It should thus be investigated over a wider range of objects.
The warm/hot (>100 K) molecular layer of the PDR is predicted to have a small spatial extent (mpc scale for conditions typical of dense galactic PDRs such as NGC~7023~NW or the Orion Bar), which is close to or below the angular resolution capability of most IR and sub-mm telescopes. Further ALMA observations and future JWST observations will thus be crucial to resolve the structure of the PDR and constrain the dynamics of photoevaporation fronts.
In addition, we find that under some conditions neutral photoevaporation flows can exist. While observational evidence of ionized flows exist, neutral flows could be searched for in PDRs that do not directly border an ionized region, through the kinematics of specific tracers of the atomic layer such as the [C\textsc{ii}] 158 $\mu$m line.

\begin{acknowledgements}
  We   acknowledge   funding   support   from   the   European Research  Council  (ERC) under grant ERC-2013-Syg-610256-NANOCOSMOS and from the Spanish
MINECO through grants AYA2012-32032 and AYA2016-75066-C2-1-P.
\end{acknowledgements}

\bibliographystyle{aa} %
\bibliography{Photoevaporation} %

%
%

\end{document}